\documentclass{aa}  
\usepackage{graphicx}
\usepackage{natbib}
\usepackage{hyperref}
\usepackage[varg]{txfonts}
\usepackage{xcolor}
\usepackage{float}

\begin{document} 

  \title{An asteroseismic age estimate of the open cluster NGC\,6866 using \textit{Kepler} and \textit{Gaia}}
  
  \author{K. Brogaard
          \inst{\ref{difa}, \ref{aarhus}}
          \and
          T. Arentoft
                    \inst{\ref{aarhus}}
                    \and
        A. Miglio \inst{\ref{difa},\ref{oas}}
                              \and
        G. Casali\inst{\ref{difa},\ref{oas}} \and
        J. S. Thomsen\inst{\ref{difa}, \ref{aarhus}} \and
        M. Tailo\inst{\ref{difa}} \and
        J. Montalb{\'a}n\inst{\ref{difa}} \and
        V. Grisoni\inst{\ref{difa},\ref{oas}} \and \\
        E. Willett\inst{\ref{bhm}} \and 
        A. Stokholm\inst{\ref{bhm}, \ref{difa}, \ref{oas}, \ref{aarhus}} \and
        F. Grundahl\inst{\ref{aarhus}}\and
        D. Stello\inst{\ref{nsw},\ref{syd},\ref{aarhus}} \and
        E. L. Sandquist\inst{\ref{sdsu}}
 }

  \institute{
  {Department of Physics \& Astronomy, University of Bologna, Via Gobetti 93/2, 40129 Bologna, Italy}\label{difa}
  \and
  {Stellar Astrophysics Centre, Department of Physics \& Astronomy, Aarhus University, Ny Munkegade 120, 8000 Aarhus C, Denmark}\label{aarhus}
  \and
{INAF – Osservatorio di Astrofisica e Scienza dello Spazio, Via P. Gobetti 93/3, 40129 Bologna, Italy} \label{oas}
\and
{School of Physics and Astronomy, University of Birmingham, Edgbaston, Birmingham, B15 2TT, UK}
\label{bhm}
\and
{School of Physics, University of New South Wales, NSW 2052, Australia}\label{nsw} 
\and
{Sydney Institute for Astronomy (SIfA), School of Physics, University of Sydney, NSW 2006, Australia}\label{syd}
\and
{Department of Astronomy, San Diego State University, San Diego, CA 92182, USA}\label{sdsu}
    }

  \date{Received XXX / Accepted XXX}

  \abstract{
   Asteroseismic investigations of solar-like oscillations in giant stars allow the derivation of their masses and radii. For members of open clusters this allows an age estimate of the cluster which should be identical to the age estimate from the colour-magnitude diagram, but independent of the uncertainties that are present for that type of analysis. Thus, a more precise and accurate age estimate can be obtained. 
}{
    We aim to identify and measure asteroseismic properties of oscillating giant members of the open cluster NGC\,6866 and utilise these for a cluster age estimate. Model comparisons also allow constraints on the stellar physics, and here we investigate the efficiency of convective-core overshoot during the main-sequence, which has a significant influence on the age for these relatively massive giants. Effects of rotation and core overshoot are similar, but not identical, and we therefore also investigate the potential of our measurements to distinguish between these effects.
}{
    We identify six giant members of NGC\,6866 via photometry, proper motions, and parallaxes from \textit{Gaia} and spectroscopic literature measurements. These are combined with asteroseismic measurements which we derive using photometric data from the \textit{Kepler} mission for five of the stars. Comparisons to stellar-model isochrones constrain  the convective-core overshoot and enables a more precise and accurate age estimate than previously possible.
}{
    A significant amount of differential reddening is found for NGC\,6866. Asteroseismology establishes the helium-core burning evolutionary phase for the giants, which have a mean mass of 2.8 $M_{\odot}$. Their radii are significantly smaller than predicted by current 1D stellar models unless the amount of convective-core overshoot on the main sequence is reduced to $\alpha_{\rm ov} \leq 0.1\cdot H_p$  in the step-overshoot description. Our measurements also suggest that rotation has affected the evolution of the stars in NGC\,6866 in a way that is consistent with 3D simulations but not with current 1D stellar models.
    The age of NGC\,6866 is estimated to be $0.43\pm0.05$ Gyr, which is significantly younger and more precise than most previous estimates. 
}{
    We derive a precise cluster age while constraining convective-core overshooting and effects of rotation in the stellar models. A comparison to age estimates from machine learning methods of the same and similar giant stars uncovers potential biases for automated asteroseismic and non-asteroseismic age estimates of helium-core burning stars. 
    }

\maketitle

\section{Introduction}

Open and globular star clusters offer the opportunity to investigate stellar evolution in detail thanks to the common properties of their member stars. Asteroseismology of solar-like oscillators is a strong tool for improving such studies further \citep[e.g.][]{Miglio2012,Miglio2016,Handberg2017,Arentoft2019,Sandquist2020,Tailo2022,Howell2022}. However, asteroseismology requires long uninterrupted high-precision observations. This currently limits detailed asteroseismic cluster studies to those observed by CoRoT \citep{Baglin2006}, \textit{Kepler} \citep{Borucki2010} or K2 \citep{Howell2014}, with few exceptions \citep[e.g. Epsilon Tau analysed by][]{Arentoft2017,Brogaard2021a}. Of these missions, CoRoT only observed one open cluster, NGC\,6633 \citep{Poretti2015,Lagarde2015}, while K2 offered only a shorter time span on each observed field and thus a reduced amount of observations \citep[e.g.][]{Stello2016,Sandquist2020} compared to the primary \textit{Kepler} mission.    

There are four open clusters in the original \textit{Kepler} field: NGC\,6791 (8.3 Gyr; \citealt{Brogaard2012,Brogaard2021b}), NGC\,6819 (2.4 Gyr; \citealt{Brewer2016}), NGC\,6811 (1.0 Gyr; \citealt{Molenda2014,Sandquist2016}), and NGC\,6866 (0.43 Gyr; this work). The three first have already been given substantial, though not complete, asteroseismic attention. The latter, NGC\,6866, has surprisingly not been studied in detail exploiting asteroseismology of solar-like oscillating members. The only literature references found for \texttt{title:(NGC6866 asteroseismology)} on NASA ADS (on 2023-02-03) were about $\delta$\,Sct stars \citep{Matobe2021} or stars that turn out to be non-members \citep{Gai2018, Tang2018}. In addition to these, \citet{Balona2013} studied pulsations in the field of NGC\,6866 including solar-like oscillations. However, their study was prior to the data releases of  \textit{Gaia} \citep{GaiaDR3-2022,Gaia2018,Gaia2016}, which complicated the membership determinations, leaving them inconclusive. As it turns out, their table 7 with stars displaying solar-like oscillations in the field of NGC\,6866 only has one star that remains a cluster member in our current study.

NGC\,6866 is a relatively young open cluster, but there is poor consensus on its precise age in the literature. Even according to relatively recent literature, it is $0.48$ \citep{Kharchenco2005}, $0.65$ \citep{Cantat2020}, or $0.78$\,Gyr \citep{Bossini2019} old. This relatively wide range of cluster ages is likely related to issues with matching isochrones to colour-magnitude diagrams (CMDs) of young open clusters, where member stars are few and scattered in the turn-off region and there are no subgiants to guide the isochrone fitting. This is made more difficult by theoretical uncertainties regarding the extent of the convective cores, which has significant impact on the age of stars in this age range \citep{Lebreton2014}.

Here, we identify and investigate oscillating giant members of the open cluster NGC\,6866 and use them to constrain cluster properties including the age. The paper outline is as follows : First, we present our identification of clusters members in Sect.~\ref{sec:observations} and derive their luminosities in Sect.~\ref{sec:luminosity}. Sect.~\ref{sec:kepler} describes our data reduction and Sect.~\ref{sec:asteroseismology} the asteroseismic analysis, leading to the derivation of masses and radii. We then carry out comparisons to stellar-model isochrones as well as the determination of the cluster age in Sect.~\ref{sec:comparisons}. Potential consequences of our results for other areas of astrophysics are discussed in Sect.~\ref{sec:results}. Our summary, conclusions, and outlook are given in Sect.~\ref{sec:conclusions}. 

\section{Identifying targets and their properties}
\label{sec:observations}
\begin{figure*}
   \centering
    \includegraphics[width=\hsize]{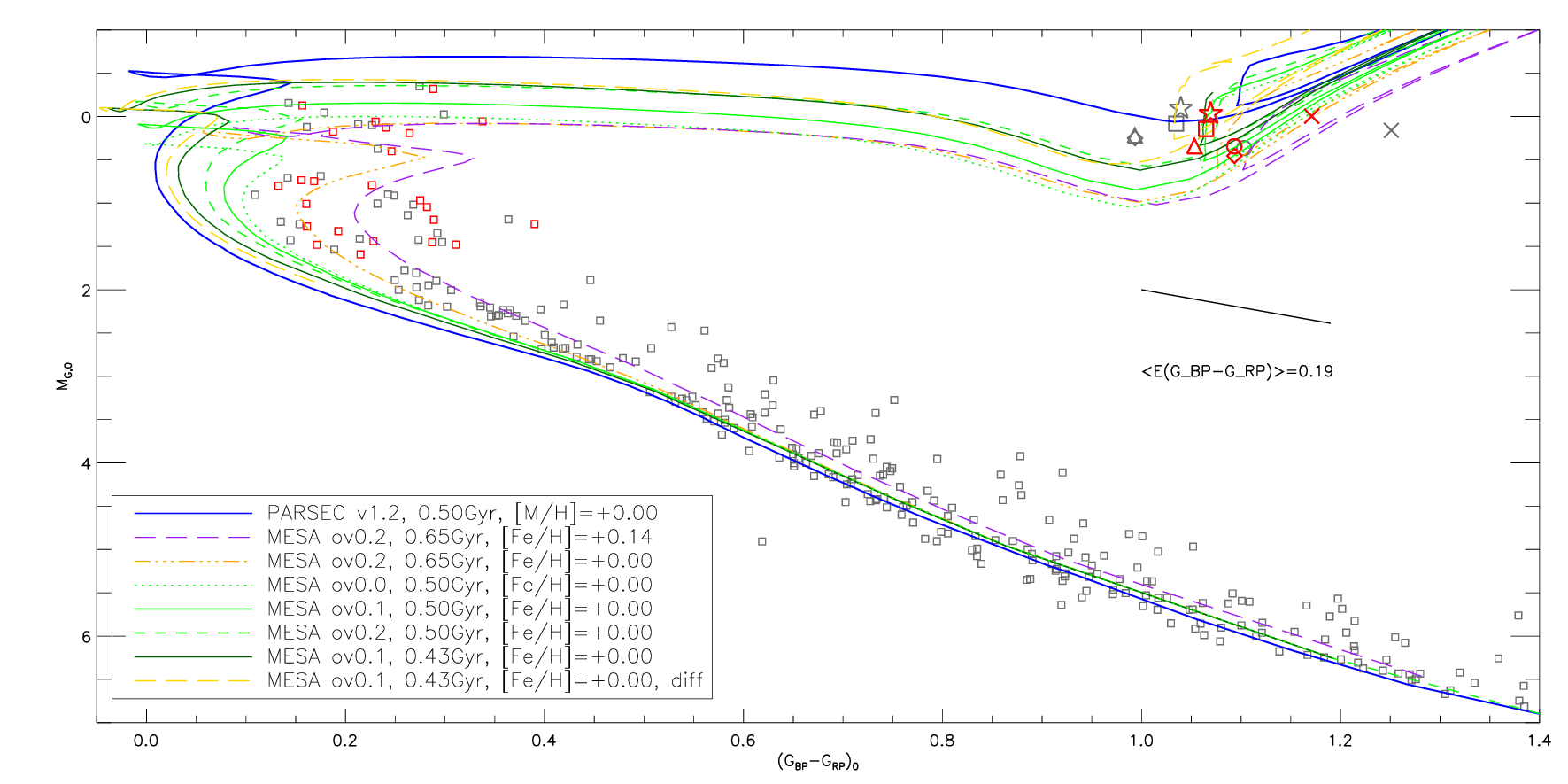}
    \includegraphics[width=9cm]{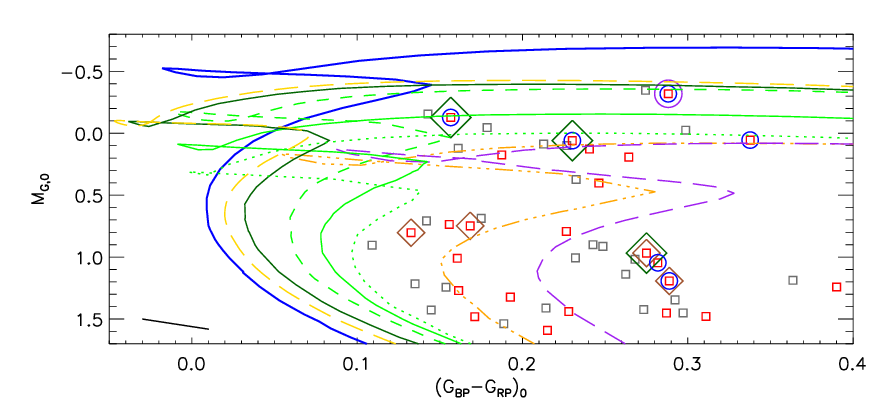}
    \includegraphics[width=9cm]{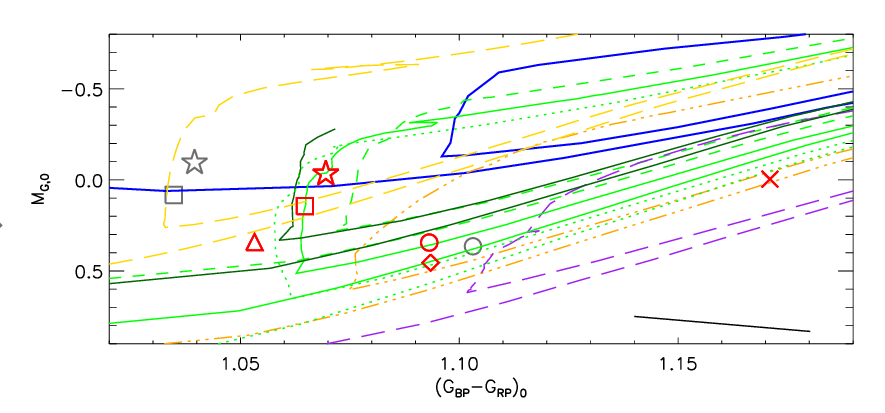}
      \caption{\textit{Gaia} colour-magnitude diagrams of NGC\,6866 proper motion and parallax members.\\
      \textit{top panel}: Grey squares mark members that have been shifted to the absolute and unreddened scale using the mean reddening for the 6 giants and 25 brightest main-sequence members. The black line shows the mean reddening vector. 
      Small red squares at the top of the main-sequence mark the brightest members when shifted according to individual reddening values.
      All giant members are marked by specific grey symbols with cross reference to Table~\ref{tab:data}. Corresponding red symbols mark the same giants where they have been shifted according to their individual reddening values instead of the mean. Also shown are isochrones with details in the text and legend.\\
      \textit{bottom left panel}: Zoom of the upper panel in the upper main-sequence area. The black line shows the typical one-sided 1$\sigma$ reddening uncertainty. The purple circle marks a known $\delta$\,Sct and binary star system \citep{Murphy2018} that is also a close neighbour to one of the red giants, KIC\,8264549. Dark green diamonds mark photometrically established fast rotators with periods from 2 to 5 days \citep{Nielsen2013}. Brown diamonds mark fast rotators with $v$sin$i$ from 130-159 $\rm km s^{-1}$ \citep{Frasca2016}. Blue circles mark stars classified as photometrically variable in \textit{Gaia} DR3.\\
      \textit{bottom right panel}: Zoom of the upper panel in the HeCB area.
      The black line shows the typical one-sided 1$\sigma$ reddening uncertainty. Symbols are as in the upper panel.
      }
         \label{fig:CMD}
   \end{figure*}

To identify giant members of NGC\,6866 we used TOPCAT \citep{Taylor2005} with the \textit{Gaia} DR3 catalogue \citep{GaiaDR3-2022}. We first selected stars in the sky region of NGC\,6866 that shared similar proper motions and parallaxes (Proper motion in right ascension direction from $-1.767$ to $-1.036$ $\rm mas\cdot yr^{-1}$, Proper motion in declination direction from $-6.078$ to $-5.425$ $\rm mas\cdot yr^{-1}$, parallax from $0.499$ to $0.935$ mas before zero-point correction). For those stars we plotted the colour-magnitude diagram (CMD) shown in Fig.~\ref{fig:CMD} and identified the individual stars in the CMD where one would expect to find cluster giants (bright cool stars). We found 6 potential cluster members this way. These were cross-matched with membership info from \citet{Cantat2020}, where all but one were listed and had a non-zero (though not all large) membership probability. Although not selected according to radial velocity, the 6 giants also turn out to have very similar line-of-sight velocities in the \textit{Gaia} data, further supporting their cluster membership. Table~\ref{tab:data} provides an overview of the stellar properties of these stars, either collected from the literature or from this work.

\begin{table*}[hbt!]
\caption{Information on giant members of NGC\,6866}  
\label{tab:data}      
\centering                          
\begin{tabular}{l c c c c c c }        
\hline\hline                 
KIC ID  & KIC8461659 & KIC8329894 & KIC8395903  &  KIC8264549 &  KIC7991875 & KIC8264592      \\
\hline                                   
Symbol in plots & cross & square & circle & triangle & star & diamond  \\
\textit{Gaia} DR3 ID & 208189159- & 207606690- & 208207250- & 207587768- & 207586586- & 207587709-\\
             & 4650263424 & 1752525440 & 2978527488 & 2673012480 & 2922829056 & 8557450112\\
2MASS ID & 20043730 & 20032282 & 20034631 & 20035516 & 20030966 & 20035769\\
         & +4425100 & +4415502 & +4422526 & +4408242 & +4346055 & +4406276\\
RA (degrees)  & 301.155417 & 300.845116 & 300.942946 & 300.979852 & 300.790240 & 300.990398 \\
Dec (degrees) & 44.419404  & 44.263915 & 44.381256 & 44.140023 & 43.768202 & 44.107644\\
$K_S$ (mag) & 8.760(16) & 9.14(16) & 9.166(14)  & 9.235(18) & 8.931(16)  & 9.351(16)   \\
$G$ (mag) & 11.287534 & 11.302372 & 11.438034 & 11.360455 & 11.102879 & 11.447467  \\
$G-K_S$ (mag) & 2.53 & 2.157 & 2.272 & 2.125 & 2.172 & 2.096 \\
$G_{\rm BP}-G_{\rm RP}$ DR3 & 1.4409981 & 1.2247219 & 1.2931328 & 1.1832037 & 1.2294884 & 1.1834383 \\
RV (km/s) & 12.3 & 12.2 & 12.6 & 14.2 & 12.7 & 13.5\\
pmra (mas$\cdot yr^{-1}$) & -1.348(12) & -1.163(14) & -1.335(13) & -1.227(33) & -1.316(17) & -1.177(13) \\
pmdec (mas$\cdot yr^{-1}$) & -5.848(12) & -5.728(13) & -5.905(13) & -5.742(32) & -5.855(16) & -6.141(13)\\
\textit{Gaia} DR3 parallax (mas) & 0.682(11) & 0.652(12) & 0.700(11) & 0.677(29) & 0.659(13) & 0.661(11)\\
$\nu$\_eff & 1.4282342 & - & 1.4545882 & 1.4765111 & 1.4668154 & 1.476384\\
pseudocolour & - & 1.4647219 & - & - & - & - \\ 
npar & 5 & 6 & 5 & 5 & 5& 5\\
ecl\_lat (deg) & 62.285942 & 62.244026 & 62.318716 & 62.090355 & 61.814700 & 62.057926 \\
RUWE & 0.8569876 & 0.99588126 & 0.85863876 & 2.0729961 & 0.90071553 & 0.842494\\
Parallax corr. (mas) & -0.02964 & -0.03021 & -0.03256 & -0.03461 & -0.03004 & -0.03490\\
Distance (pc) & 1405 & 1466 & 1365 & 1405 & 1451 & 1437 \\
$(m-M)_0$ (mag) & 10.74 & 10.83 & 10.68 & 10.74 & 10.81 & 10.79 \\
P\_mem\tablefootmark{a}  & 0.4 & 0.9 & 0.7 & 0.6 & - & 0.3\\
\hline
$T_{\rm eff}$ (K)\tablefootmark{b} & 5044 & 5114 & 5104 & 5201 & - & -\\
$T_{\rm eff}$ (K)\tablefootmark{c} & 5045 & - & - & - & - & -\\
$T_{\rm eff}$ (K)\tablefootmark{d} & 5032 & 5069 & - & - & - & -\\
$T_{\rm eff}$ (K)\tablefootmark{e} & 4983 & - & 5052 & - & 4969 & -\\
$T_{\rm eff}$ (K)\tablefootmark{f} & 4998 & 5052 & 5052 & 5052 & - & 5037\\
$T_{\rm eff}$ (K)\tablefootmark{g} & - & 5092 & 5106 & 5126 & -  & -\\
$T_{\rm eff}$ from DR3 $G_{\rm BP}-G_{\rm RP}$ (K)\tablefootmark{h} & 4864 & 5091 & 5037 & 5106 & 5078  & 4975 \\
$T_{\rm eff}$ from $G-K_S$ (K)\tablefootmark{i} & 4978 & 5157 & 5120 & 5112 & 5137 & 5037 \\
$T_{\rm eff}$ (K)\tablefootmark{j} & 5044 & 5096 & 5101 & 5117 & -  & 5105\\
$\rm [M/H]$\tablefootmark{j} & $+0.021$ & $+0.027$ & $-0.011$ & $+0.021$ & - & $+0.021$\\
$\rm [C/N]$\tablefootmark{j} & $-0.636$ & $-0.522$ & $-0.535$ & $-0.608$ & - & $-0.561$\\
\hline
$E(B-V)$ calculated \tablefootmark{k} & 0.294 & 0.139 & 0.193 & 0.132 & 0.127 & 0.114\\
$E(B-V)$ Green \tablefootmark{l} & $0.27_{-0.03}^{+0.03}$ & $0.16_{-0.02}^{+0.03}$ & $0.20_{-0.03}^{+0.02}$  & $0.13_{-0.02}^{+0.03}$ & $0.16_{-0.02}^{+0.03}$ & $0.09_{-0.01}^{+0.02}$\\
$E(B-V)$ from NaD-lines \tablefootmark{m} & - & 0.12(2) & 0.24(3)  & - & - & 0.12(1)\\
$E(B-V)$ Lallement \tablefootmark{n} & $0.173(35)$ & $0.166(36)$ & $0.161(34)$  & $0.156(34)$ & $0.160(34)$ & $0.160(33)$\\
BC$_G$ (mag) & -0.020 & 0.012 & 0.007 & 0.012 & 0.006 & -0.004 \\
BC$_{K_S}$ (mag)& 1.933  & 1.841 & 1.857 & 1.841 & 1.861 & 1.891\\
$L(L_{\odot})$ from $K_S$ & 94.4(3.5) & 70.4(2.9) & 61.1(2.2) &  60.1(5.1) &  81.3(3.5) &  55.7(2.1)\\
$R(R_{\odot})$ SB & 12.72(56)   &    10.76(47)   &    10.01(43)  &     9.86(57)  &     11.81(53)  &  9.55(41)\\
\hline
\end{tabular}
\tablefoot{  
\tablefoottext{a}{\citet{Cantat2020}.}
\tablefoottext{b}{LAMOST DR5, \citet{Zhong2020}.} 
\tablefoottext{c}{APOKASC, \citet{Pinsonneault2018}.} 
\tablefoottext{d}{\citet{Yu2018}.}
\tablefoottext{e}{LAMOST, \citet{Frasca2016}.}
\tablefoottext{f}{\citet{Ness2016}.}
\tablefoottext{g}{\citet{Carrera2019}.}
\tablefoottext{h}{Adopting $E(B-V)$ and requiring bolometric corrections from \citet{Casagrande2018} to reproduce observed colour. [Fe/H] $= +0.0$ and log$g$ = 2.7 was assumed. A change of $\pm0.1$ dex in [Fe/H] corresponds to $\pm$17 K, a change of $\pm0.1$ in log$g$ yields $\pm3$ K, and $\pm0.03$ in $E(B-V)$ gives about $\pm$80 K.}
\tablefoottext{i}{as $(h)$, but also using \citet{Casagrande2014}.}
\tablefoottext{j}{APOGEE DR17. Taken from https://skyserver.sdss.org/dr18/SearchTools/IRQS.}
\tablefoottext{k}{Calculated by assuming APOGEE $T_{\rm eff}$s from (j) and requiring bolometric corrections from \citet{Casagrande2018,Casagrande2014} to yield the observed $G-K_S$.} 
\tablefoottext{l}{From Bayestar2019 $E(g-r)$, assuming a mean of the two conversion factors 0.884 and 0.996 given in \citet{Green2019}. $A_G = 2.74\times E(B-V)$ \citet{Casagrande2018}. $A_{K_S} = 0.366\times E(B-V)$ \citet{Casagrande2014}.}
\tablefoottext{m}{Using \citet{Munari1997} and spectra from FIES at the Nordic Optical Telescope.}
\tablefoottext{n}{Using \citet{Lallement2018}.}
}
\end{table*}

We searched the literature for spectroscopic $T_{\rm eff}$ measurements of the 6 targets, which we list in Table~\ref{tab:data}. Not all targets were measured in one single study, so we first made an attempt to put all the six stars on the same spectroscopic $T_{\rm eff}$ scale by exploiting overlaps of some targets between studies. We then realised that all but one of the stars were in APOGEE DR17 \citep{APOGEEDR17-2022}.

Furthermore, we derived photometric $T_{\rm eff}$ values by obtaining the reddening of each target from \citet{Green2019} and requiring $G_{\rm BP}-G_{\rm RP}$ or $G-K_S$ colours to match with the predictions from the bolometric corrections (BCs) of \citet{Casagrande2014,Casagrande2018}. The $T_{\rm eff}$ values from the two colours agree within 66 K or less for four stars, and within 114 K or less for the other two with the tendency that $G-K_S$ always produces the higher values compared to $G_{\rm BP}-G_{\rm RP}$. The general agreement is as would be expected when taking uncertainties in photometry, reddening and colour-BC relations into account. The agreement between the photometric and spectroscopic $T_{\rm eff}$ values is also better than 100 K in all but a few cases. 

All these estimates are given in Table~\ref{tab:data}. The close similarity of $T_{\rm eff}$ from star to star is in line with expectation from the CMD where the giants align much better in colour when accounting for the star-to-star differences in reddening, as can be seen in Fig.~\ref{fig:CMD}. The stellar-model isochrones predict this behaviour for the helium-core burning (HeCB) phase, where these stars are expected to be (cf. Sect.~\ref{sec:evol}). However, the uncertainties in reddenings are at the $\pm0.03$ level and if we had taken the $E(B-V)$ estimates from \citet{Lallement2018} instead, we would not have obtained such a nice alignment. Alternatively, we could have obtained an even better star-to-star $T_{\rm eff}$ agreement, as predicted by the models, if we had adjusted the reddening estimates further within the measurement uncertainties. This could potentially be improved by measuring reddening via the interstellar Na D lines \citet{Munari1997}, if we had spectra of all the targets in that wavelength region, which unfortunately we did not. We did find archival optical spectra from the FIES spectrograph at the Nordic Optical Telescope for three of the stars and determined their $E(B-V)$ values following the procedures described in \citet{Brogaard2011,Sandquist2016}. Basically, one measures the equivalent-width (EW) of the interstellar Na D lines and uses the \citet{Munari1997} calibration to translate to reddening. We estimated the uncertainty as half the difference obtained from each of the two Na D lines individually. The relative star-to-star differences are likely smaller than this since the same line gives the higher value for all stars, and thus the uncertainty is mainly systematic. One star had three individual spectra and the spectrum-to-spectrum differences in derived $E(B-V)$ was only 0.003, indicating that the star-to-star differences should be very precise. However, the Na D lines in the spectra of the NGC\,6866 giants show multiple components in which case one can only use the reddening values as upper limits according to \citet{Munari1997}. We give all our reddening estimates in Table~\ref{tab:data}. The Na lines seem to support the larger star-to-star $E(B-V)$ variation suggested by \citet{Green2019} compared to \citet{Lallement2018}. If one adopts the APOGEE DR17 spectroscopic $T_{\rm eff}$, $E(B-V)$ can be calculated by requiring a match to the photometric temperatures. Such results, shown in Table~\ref{tab:data}, also agree well with the \citet{Green2019} 3D reddening map.

\subsection{Spectroscopic analysis of KIC\,8395903}

For one of the targets, KIC\,8395903, we combined two high-resolution ($R \sim 67,000$) FIES \citep{Telting2014} spectra\footnote{Filenames FIwh150083.fits and FIwh150084.fits} from the archive at the Nordic Optical Telescope with a signal-to-noise ($S/N$) level high enough to allow derivation of spectroscopic metallicity and effective temperature (the combined spectrum has a $S/N$ $\sim$ 77 at 6500$\AA$). 
We performed a spectral analysis based on EW measurements. The EWs of the spectral absorption lines were measured using DOOp \citep[Daospec Output Optimiser pipeline,][]{Cantat2014}, an automated wrapper of DAOSPEC \citep{stetson08}. The line-list was prepared for the spectral analysis of the \textit{Gaia}-ESO survey \citep{heiter15a}. Finally, the atmospheric parameters were determined with MOOG in the automatic form using FAMA \citep[Fast Automatic MOOG Analysis,][]{Magrini2013}. Specifically, the code searches for the three equilibria in an iterative way (excitation, ionisation, and the trend between the iron abundances and the reduced EW, $\log[EW/\lambda]$). The iterations are executed with a series of steps starting from a set of initial parameters and arriving at the final set of parameters that simultaneously fulfil the equilibria. FAMA uses MOOG in its 2017 version \citep{sneden12} and MARCS model atmospheres \citep{Gustafsson2008}. To compute the Solar-scaled metallicity [Fe/H], we define our Solar scale measuring the iron abundance on a solar spectrum. For this task we use a spectrum of Vesta collected by the twin HARPS spectrograph at the 3.6 m ESO telescope. We get an iron abundance for the solar spectrum of $7.50\pm0.07$ dex in agreement with \citet{Asplund2009}. Using this solar iron abundance, we obtain for KIC\,8395903 the solar-scaled metallicity [Fe/H] shown in Table~\ref{tab:specpar}.

We derived the stellar parameters and metallicity with and without fixing $\log g$ to the asteroseismic value determined later. As seen, our measured $T_{\rm eff}$ is higher than all spectroscopic and photometric values given in Table~\ref{tab:data} for this star and all the other HeCB stars, although compatible within $1\sigma$ mutual uncertainties. Because we are only able to derive $T_{\rm eff}$ for one of the giants, we rely instead on the APOGEE DR17 values in the following analysis. A simple increase of all APOGEE $T_{\rm eff}$ values to the temperature scale suggested by our spectral analysis of this one star, which is about 160 K hotter, would make the self-consistency among asteroseismic mass equations worse for all but one star in Sect.~\ref{sec:masses} though they remain in agreement within mutual 1$\sigma$ uncertainties in Table~\ref{tab:seispropdata}. Increasing the $T_{\rm eff}$ values by another 100 K destroys the mass self-consistency for all stars so that for each star the masses from the four equations are no longer consistent within their mutual 1$\sigma$ uncertainties. The same happens if one decreases the temperatures 200 K below APOGEE $T_{\rm eff}$ values. While this suggests that the values we derive are close to the correct $T_{\rm eff}$ values, both statistical and systematic uncertainties are unfortunately still relatively large.
For [Fe/H] we find KIC\,8395903 to be slightly super-solar with the solar value within the uncertainty. This is in agreement with APOGEE DR17 where [Fe/H]=$-0.023$ and [M/H]=$-0.01$ for this particular star, while [Fe/H] is between $+0.005$ and $+0.015$ and [M/H] very close to $+0.02$ for the other four HeCB stars measured; the exact numbers are given in Table~\ref{tab:data}. In later sections we will therefore be comparing to isochrones that assume either a solar metallicity or a slightly super-solar value.

\begin{table}
\caption{Spectroscopic parameters of KIC\,8395903}  
\label{tab:specpar}      
\begin{tabular}{l c c c }        
\hline\hline                 
Fixed log$g$ & log$g$ (cgs) & $T_{\rm eff}$ (K) & [Fe/H]    \\
\hline
No & 2.99(25) & 5184(114) & +0.00(9) \\
Yes & 2.86 & 5260(106) & +0.04(9) \\
\hline                                   
\end{tabular}

\end{table}

\subsection{Adopted effective temperatures}

Our investigations of the effective temperatures in the preceding sections show that the photometric and spectroscopic $T_{\rm eff}$ values are generally in good agreement. The same is true among different spectroscopic studies, albeit with differences at the level of about 100 K. In the end, we decided to adopt the APOGEE DR17 $T_{\rm eff}$ values and an uncertainty of 100 K for the asteroseismic analysis, keeping an eye out for potential biases this might cause. A future spectroscopic study with high $S/N$ and high-resolution optical spectra could potentially improve on the precision and accuracy of the $T_{\rm eff}$ and metallicity, especially if done in a differential way with respect to other open clusters at near-solar metallicity \citep{Slumstrup2019} and combined with measurements of the interstellar Na D lines for reddening estimates \citep{Munari1997}.

\section{Luminosities}
\label{sec:luminosity}
Luminosity estimates were derived for the six giant stars by combining the \textit{Gaia} DR3 parallaxes with photometry. 
The parallaxes were zero-point corrected following \citet{Lindegren2021} with all the parameters needed taken from the \textit{Gaia} archive and given in Table~\ref{tab:data} along with the derived correction. Since the stars and their sky locations are quite similar, so are the parallax corrections. The \textit{Gaia} DR3 parallaxes with zero-point corrections by \citet{Lindegren2021} were shown by \citet{Khan2023} to be in excellent agreement with asteroseismic predictions for stars in the \textit{Kepler} field, and we thus expect the same to hold true for our stars in NGC\,6866, which are located there. 

As seen in Table~\ref{tab:data}, the parallax uncertainties are from 0.011 to 0.013 mas except for KIC\,8264549, which has an uncertainty of 0.032 mas. Although this star also has a high RUWE value above 2.0, which could indicate e.g. binarity or other biases, the parallax value is in very good agreement with the other members. The mean parallax (after zero-point correction) of the six giants is $0.704\pm0.017$ mas and the parallax for KIC\,8264549 is closer to this mean than any of the other stars despite the larger uncertainty stated. This suggests that the potential bias flagged by the larger uncertainty and the high RUWE value is not present. The rms of the parallaxes of the 6 stars (0.017 mas) is slightly larger than suggested by the uncertainty for the individual targets. Although studies in the literature \citep[e.g.][]{MazApellaniz2021} have shown that the \textit{Gaia} parallax uncertainties are underestimated we found another potential explanation; assuming that the stars differ in position by about the same amount in the direction along the line of sight as they do perpendicular to it, we calculated that the stars could have real parallax differences of the order of 0.01 mas. If such a contribution of 0.01 mas is added in quadrature to the $0.011-0.013$ mas uncertainty, then there is agreement with the rms across the stars, and thus no additional error contribution is needed. This does, however, not mean that no error is present. If relying on \citet[e.g.][]{MazApellaniz2021} to calculate external parallax uncertainties they should be about 0.0184 for our stars, and thus would leave no room for true positional variations in the radial direction unless much of the external uncertainty is considered systematic.  
We notice that the two stars that have the smallest and the largest parallax, respectively, are also those that have the largest inconsistencies in the asteroseismic mass analysis later. However, due to uncertainties on both the parallaxes and the asteroseismic parameters, we are not able to conclude whether adopting the mean parallax for all stars would be more accurate.

We adopted the APOGEE DR17 spectroscopic temperatures and metallicities and derived reddening estimates by requiring photometric $T_{\mathrm{eff}}$ values from $G-K_S$ colours and \citet{Casagrande2018} to agree. For KIC\,7991875, which does not have an APOGEE DR17 measurement, we adopted 5044 K, the value for KIC\,8461659, which is the target that most closely resembles KIC\,7991875 in the CMD. To minimise the effect of reddening/absorption uncertainties, we used 2MASS \citep{Cutri2003} $K_S$ apparent magnitudes and $K_S$ bolometric corrections from \citet{Casagrande2014} with the parallaxes to estimate the luminosities; $A_{K_S}=0.366\times E(B-V)$, in comparison to e.g. $A_{V}=3.1\times E(B-V)$ or $A_{G}=2.74\times E(B-V)$ and thus the effect of reddening uncertainty is minimised in the near-infrared because the coefficient is much smaller. The numbers are given in Table~\ref{tab:data}. As seen, the reddening estimates are in very good agreement with those from the \citet{Green2019} 3D reddening maps.

\section{\textit{Kepler} observations and data reduction}
\label{sec:kepler}

We used the Python programme Lightkurve \citep{Lightkurve2018} to search for \textit{Kepler} \citep{Borucki2010} light curves of the targets. The \textit{Kepler} data are divided into 90-day quarters because the \textit{Kepler} spacecraft rotated by 90 degrees every 90 days to keep the solar panels pointing towards the Sun. We found that KIC\,7991875 was not observed at all, but still kept it in Table~\ref{tab:data} so that all giant cluster members are included. KIC\,8461659 was observed for quarters 0-3, 5-8, 10-12, and 14-15, while the remaining four stars were observed in all quarters 0-17. When investigating the light curves, we found that for some of them, there were problems with contamination from neighbouring stars, while for others the standard pipeline pixel masks were just not optimal for some quarters. These issues reduced the quality of the light curves, and is likely the reason why only two of the five NGC\,6866 giant member stars observed by \textit{Kepler} have asteroseismic measurements in the literature (see Table~\ref{tab:seisdata}). To minimise these problems,
we extracted new light curves from the \textit{Kepler} pixel data with our own custom masks created manually to minimise contamination while still retaining as much target flux as possible.

With our custom masks we extracted light curves for the individual quarters. We then stitched them together using our own code inspired by \citet{Handberg2014} to adjust the quarter-to-quarter flux-level variations. The light curves were filtered using a moving median with a width of 3 days, and  outliers above 3.5$\sigma$ were removed. In appendix A, we provide information on potential contaminants that we attempted to avoid for the individual targets. 

\section{Asteroseismology}
\label{sec:asteroseismology}

The power spectra based on the \textit{Kepler} light-curves were analysed to determine asteroseismic parameters as described in \citet{Brogaard2021c} and references therein, with only one notable difference, which is the treatment of the stellar background when determining the frequency of maximum power, $\nu_{\rm max}$; in the present case we apply a combination of a linear fit and a Gaussian envelope as described in \citet{Mosser2009} and applied in \citet{Arentoft2019}, instead of the more complicated formula using Harvey models described in \citet{Handberg2017} and applied in \citet{Arentoft2017} and \citet{Brogaard2021c}. The reason for this is that we, for the present paper, have applied a different method for filtering out low-frequency variations in the \textit{Kepler} light-curves than for the two latter papers, which meant that the fit with the Harvey models turned out to be unstable. Instead we used the more robust fitting form with fewer free parameters as applied successfully by \citet{Mosser2009}. We tested as described below that this procedure did not bias our $\nu_{\rm max}$ measurements, and could furthermore compare our results to literature values for two of the five stars and found very good agreement. Apart from this, the analysis follows the methods described in \citet{Brogaard2021c}.

\begin{figure*}
  \centering
    \includegraphics[width=\hsize]{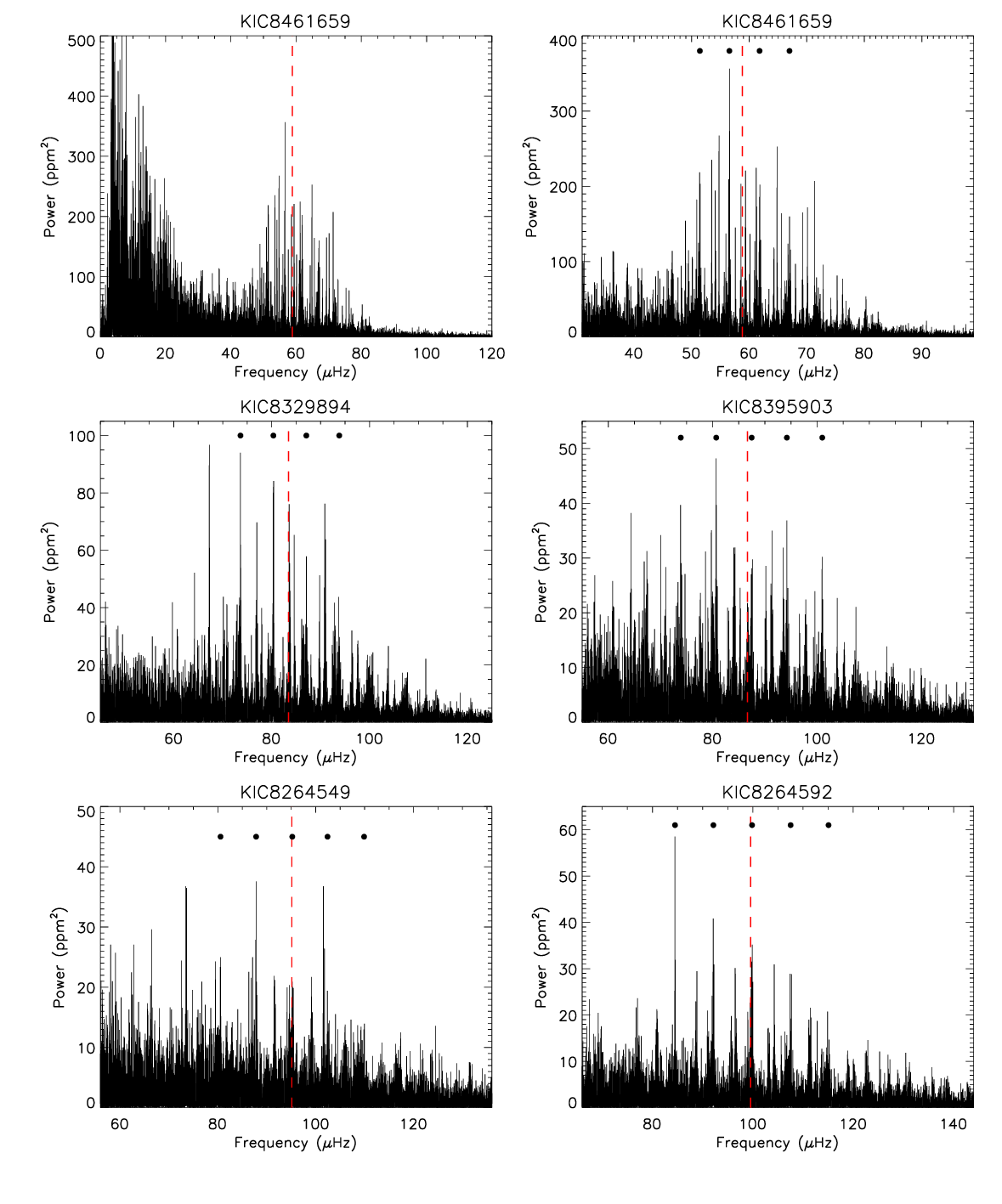}
      \caption{Power spectra of the five target stars, with the spectrum of one of them, KIC8461659, shown twice in the upper panels: one version (upper left) which extends to low frequencies, and one version (upper right) focusing on the frequency region of the oscillations. The effect of the filtering in the preparation of the light curve can be seen in the leftmost version of these, where the power drops off at low frequencies. We only show the version focused on the oscillations for the remaining four stars. The frequency of maximum power, $\nu_{\rm max}$, is indicated in each panel as a (red) vertical dashed line. The filled circles indicate the position of the 4-5 dominating $\ell=0$-modes near $\nu_{\rm max}$. Due to the stochastic nature of the oscillations, each mode is represented by a series of close peaks, which is the reason why the detailed frequency analysis is performed using smoothed versions of the power spectra (see text).}
         \label{fig:PS}
   \end{figure*}

\begin{figure}
   \centering
    \includegraphics[width=\hsize]{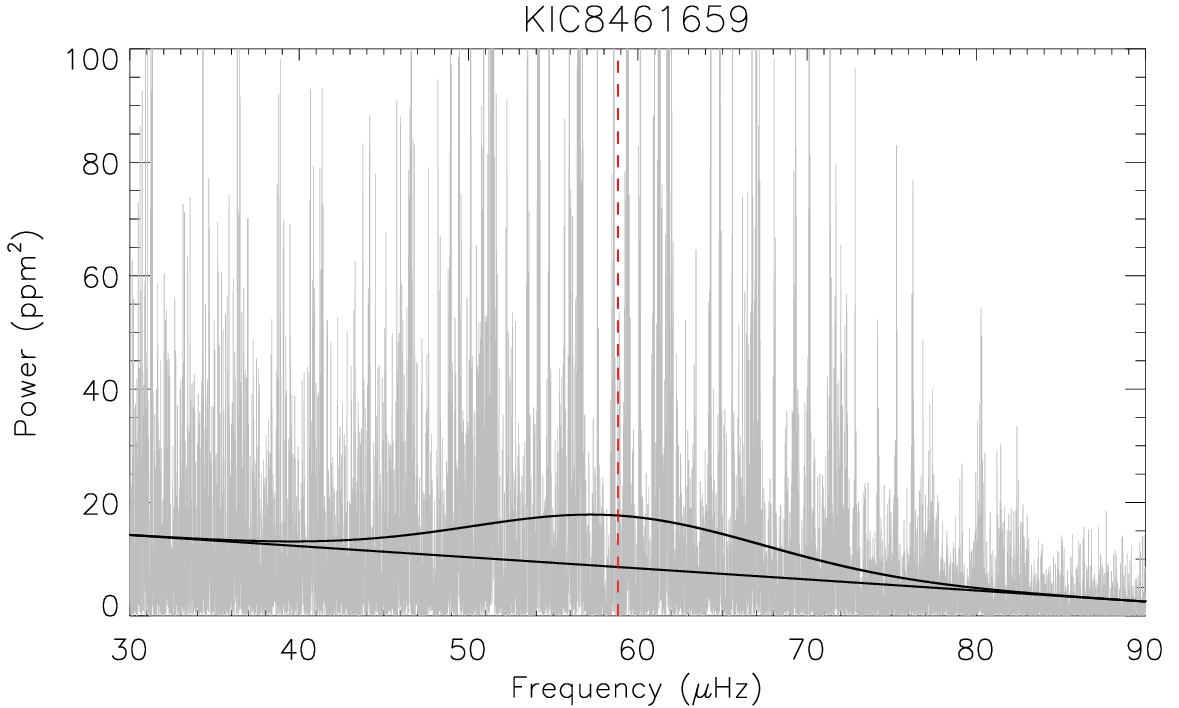}
      \caption{The frequency of maximum power, $\nu_{\rm max}$ is determined from a model including a linear trend and a Gaussian envelope. The linear trend accounts for the stellar background in the frequency region of the oscillations, which increases towards lower frequencies, while the Gaussian envelope fits the solar-like oscillations which follows a Gaussian distribution centered on $\nu_{\rm max}$. The determined value for $\nu_{\rm max}$ for the star shown here as an example, KIC\,8461659, is indicated by the vertical (red) dashed line, and the underlying, fitted power spectrum is shown in grey.}
         \label{fig:Fit}
   \end{figure}

The analysed power spectra are shown in Fig.~\ref{fig:PS} and the combined fit of a linear trend and a Gaussian envelope is illustrated for one of the stars in Fig.~\ref{fig:Fit}. From such fits we determine $\nu_{\rm max}$ for each of the five target stars, resulting in the values marked in Fig.~\ref{fig:PS} and quoted in Table~\ref{tab:seisdata}. In the same process, we obtain for each star a version of the power spectrum where the stellar background is subtracted in the frequency range where the oscillations are found. We tested on a heavily smoothed version of the oscillation signal in the background-corrected spectra that the maximum of the smoothed signal coincided with the value for $\nu_{\rm max}$ determined above. After this, we used the corrected spectra for determining average large frequency separations between modes of consecutive radial order, $\Delta\nu_{\rm ps}$, by applying the same methods as in \citet{Brogaard2021c}. These values are refined later in the analysis, where we use individual, radial oscillation modes to determine $\Delta\nu_{0}$, which are the values we use in the analysis in the subsequent sections. We can however compare our results for $\nu_{\rm max}$ and $\Delta\nu_{\rm ps}$ to literature values for KIC\,8461659 and KIC\,8329894, which are quoted at the bottom of Table~\ref{tab:seisdata}. We find very good agreement between our values and the literature values, which has also been the case in our previous papers, see for example \citet{Brogaard2021c}.

\begin{figure*}
   \centering
    \includegraphics[width=\hsize]{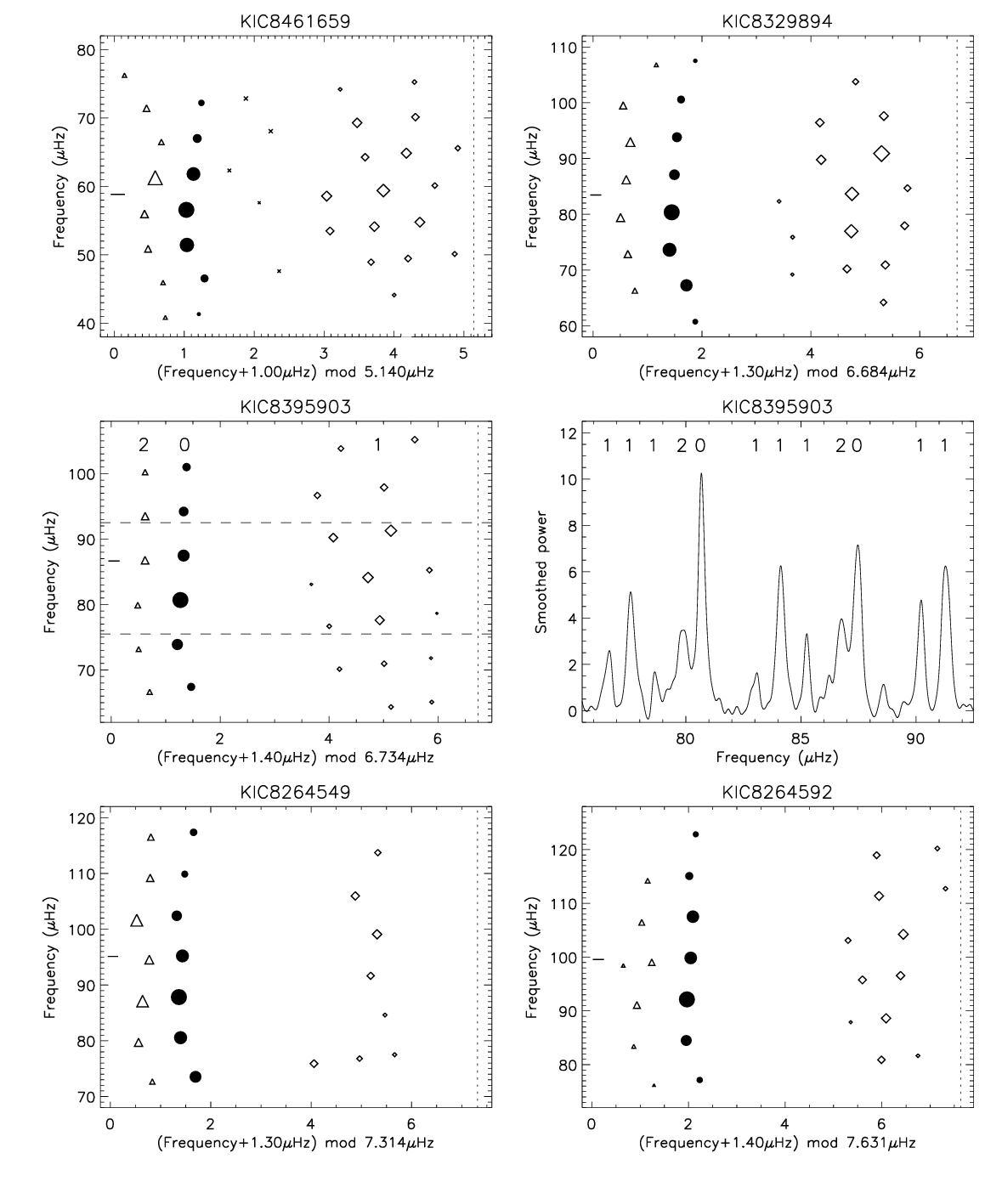}
      \caption{The top- and bottom panels along with the leftmost middle panel display the {\'e}chelle diagrams for our five stars. These diagrams show the oscillation frequencies as a function of the same frequencies modulo the large frequency separation, $\Delta\nu_{\rm 0}$, which separates modes of different $\ell$-value and hence allows for mode identification. The $x$-axis has in each case been shifted by a constant to place modes of $\ell=0,2$ on the left side of the diagrams and modes of $\ell=1$ on the right side. Modes of $\ell=0$ are shown as filled circles, $\ell=1$ as diamonds, $\ell=2$ as triangles, and possible $\ell=3$-modes are shown as crosses (upper-left panel only). The frequencies of maximum power ($\nu_{\rm max}$) are indicated as short horizontal lines near the $y$-axes and the large frequency separations ($\Delta\nu_{0}$) as vertical dashed lines. The middle panels illustrate the relation between the {\'e}chelle diagram and the power spectrum: in the {\'e}chelle diagram for KIC8395903 (middle-left) the horizontal dashed lines indicate the frequency region of the power spectrum shown in the middle-right panel, with the $\ell$-values of the oscillation frequencies written at the top of each panel.}
         \label{fig:echelle}
   \end{figure*}

The next part of the analysis is to use a smoothed version of the power spectrum to determine the individual frequencies together with their uncertainties and $S/N$ values, by applying the methods described in sect. 4 in \citet{Arentoft2017}. Briefly, the frequencies were determined from the smoothed power spectra by applying a Gaussian fit to the top part of each peak found in the region of the oscillations, while the $S/N$-values were determined from the corresponding amplitude spectrum by taking the ratio of the height in amplitude of the oscillation peaks and the median height in amplitude of the noise peaks in the frequency regions surrounding the oscillations. The uncertainties on the frequencies are determined from the standard deviation of the frequency values determined from the full time-series and the two half series, as detailed in \citet{Arentoft2017}. We detected between 22 and 37 individual frequencies in the power spectra of the five stars, and using the large frequency separation determined above, the frequency values were plotted modulo $\Delta\nu_{\rm ps}$ to construct {\'e}chelle-diagrams, which separates frequencies of different $\ell$-values and allows us to identify them as modes of $\ell=0, 1$ and 2, and for one star possibly $\ell=3$. The {\'e}chelle diagrams for the five stars and an illustration of the relation between the {\'e}chelle diagram and the smoothed power spectrum are shown in Fig.~\ref{fig:echelle}, however for these final {\'e}chelle-diagrams we have used the refined large frequency separation, $\Delta\nu_{\rm 0}$ determined below. For each of the five stars we detect several orders of $\ell=0,2$ as well as a number of $\ell=1$ modes, and we see mode splitting for the $\ell=1$-modes for four of the stars, allowing us to determine period spacings as reported below. For one star, KIC8264592, we see splitting of one of the $\ell=2$ modes, which we have also seen in similar stars in NGC\,6811 \citep{Arentoft2017}. 

Following the mode detection and identification, the $\ell=0$ modes are used for refining the value of the large frequency separation to determine $\Delta\nu_0$, using a weighted fit to the frequencies of the identified $\ell=0$ modes as a function of their radial order, as described in \citet{White2011} and applied (by us) in \citet{Arentoft2017,Arentoft2019}. In the same process, we determine the dimensionless parameter $\epsilon$ and, from the differences in frequency between neighbouring $\ell=0$- and $\ell=2$ modes, we determine the small frequency separation, $\delta\nu_{\rm 02}$. All these values are listed for the five target stars in Table~\ref{tab:seisdata}. We also determine the large frequency separation and $\epsilon$ using only the three $\ell=0$ oscillation modes closest to $\nu_{\rm max}$, $\Delta\nu_{\rm c}$ and $\epsilon_{\rm c}$, along with the ratio of the large and small frequency separations, as listed in Table~\ref{tab:seisdata}. Following \citet{Kallinger2012}, these values can be used together with the observed period spacing, $\Delta P_{\rm obs}$ to determine the evolutionary status of oscillating red giants. We therefore finally determined this latter parameter, the observed period spacing, using the method detailed in \citet{Arentoft2017} and illustrated in their fig.~4. It was possible to determine $\Delta P_{\rm obs}$ for four of the stars, while for one star, KIC8264549, we do not find evidence of a clear set of $\ell=1$ modes to determine this value, see the lower left panel of Fig.\ref{fig:echelle}. There is however a single order of $\ell=1$ modes at the bottom of the {\'e}chelle diagram for KIC8264549 which can be used for determining an estimate of the observed period spacing. For another of the stars, KIC8395903, we had enough information to also determine the asymptotic period spacing, $\Delta \Pi_{\rm 1}$, again following the methods described in Arentoft et al. (2017). The period spacings are listed in Table~{\ref{tab:seisdata}}.

We finally applied the same methods as in sect.~8 in \citet{Arentoft2017}, to investigate whether we see evidence of mode suppression \citep{Mosser2012} in the oscillation spectra; the result of this analysis is that mode suppression does not seem to be present in the five giants in NGC\,6866.

\subsection{Evolutionary states}
\label{sec:evol}
We compared our results to fig. 4 of \citet{Kallinger2012} in order to establish the evolutionary states of the stars. Our measured $\Delta P_{\rm obs}$ values between 100 and 204 $\mu$Hz in combination with $\Delta\nu_{\rm c}$ values between 5.2 and 7.7 $\mu$Hz strongly indicate that the stars are in the HeCB phase of evolution, burning He in their cores. This is supported by the $\epsilon_c$ values. For KIC\,8395903 we also measured the asymptotic period spacing $\Delta\Pi_1$ and its value is very close to model predictions for the HeCB phase in Sect.~\ref{sec:comparisons}.
Only for KIC\,8264549 the evolutionary state remains slightly inconclusive; for this star we could only measure a rather uncertain value of $\Delta P_{\rm obs}$ and although its $\epsilon_c$ falls in the region indicating the HeCB phase (see upper panel of fig. 4 in \citet{Kallinger2012}), it is also compatible with the red giant branch (RGB) phase within 1$\sigma$ unlike the other stars. However, given the close similarity of both classical and asteroseismic parameters of KIC\,8264549 and the other HeCB stars, and considering the much longer timescale of the HeCB phase compared to the RGB phase, we find it most likely that this star is also be in the HeCB phase of evolution. This is corroborated by the estimate of the observed period spacing discussed above and quoted in Table~{\ref{tab:seisdata}}. Similar considerations applies to KIC\,7991875, the giant member which was not observed by \textit{Kepler}; its CMD position and estimated luminosity and radius is strongly suggesting a HeCB star. In Sect.~\ref{sec:masses} we argue further for the five stars with asteroseismology being in the HeCB phase based on comparing the ensemble to model predictions.  

\subsection{Masses and radii of the giants}
\label{sec:masses}
We used the asteroseismic scaling relations as in previous works \citep[see e.g.][]{Brogaard2021c} to estimate the masses and radii of the giants. Thus, the radius of a star is given as 

\begin{eqnarray}\label{eq:03}
\frac{R}{\mathrm{R}_\odot} & = & \left(\frac{\nu _{\mathrm{max}}}{f_{\nu _{\mathrm{max}}}\nu _{\mathrm{max,}\odot}}\right) \left(\frac{\Delta \nu}{f_{\Delta \nu}\Delta \nu _{\odot}}\right)^{-2} \left(\frac{T_{\mathrm{eff}}}{T_{\mathrm{eff,}\odot}}\right)^{1/2}
\label{eq:01}
\end{eqnarray} with the solar reference values adopted as in \citet{Handberg2017}, $\Delta \nu _{\odot}=135.1 \mu$Hz, 
$\nu _{\mathrm{max,}\odot}=3090 \mu$Hz, and $T_{\mathrm{eff,}\odot}$ = 5772 K \citep{Prsa2016}. Corrections to $\Delta \nu$, $f_{\Delta \nu}$, were determined as in \citet{Rodrigues2017} comparing to isochrones in diagrams such as Fig.~\ref{fig:ddeltanu} while $f_{\nu _{\mathrm{max}}}=1$ was assumed due to the lack of observational evidence against it. Based on Fig.~\ref{fig:deltaP}, which closely resembles Fig.~\ref{fig:ddeltanu} although the y-axis shows the period spacing instead of $f_{\Delta \nu}$, we argue that the predicted $f_{\Delta \nu}$ must be close to the true values, given that the same structural differences give rise to the pattern for both parameters. In Fig.~\ref{fig:deltaP} we show the observed period spacing for all stars and the asymptotic one for the one star where it could be measured. The observed period spacings seem to follow a linear relation as a function of $\nu_{\rm max}$. If all the observed period spacings are shifted by the difference between the observed and asymptotic period spacing for the one star with both measurement, this line is seen to follow the isochrone predictions for the asymptotic period spacing quite well, although not perfectly. This also support the HeCB phase for the full ensemble; although it can be seen in Fig.~\ref{fig:deltaP} that the stars with the largest $\nu_{\rm max}$ values are individually also consistent with the RGB phase of some of the isochrones, it would be unlikely that that any star would be observed in the RGB phase at a $\nu_{\rm max}$ that aligns so well with the HeCB phase.  

Using our luminosity estimates (cf. Sect.~\ref{sec:luminosity}), we calculated an asteroseismic mass estimate for each star in four different ways as it has also been done in e.g. \citet{Miglio2012,Brogaard2021c}:

\begin{eqnarray}
\label{eq:02}
\frac{M}{\mathrm{M}_\odot} & = & \left(\frac{\nu _{\mathrm{max}}}{f_{\nu _{\mathrm{max}}}\nu _{\mathrm{max,}\odot}}\right)^3 \left(\frac{\Delta \nu}{f_{\Delta \nu}\Delta \nu _{\odot}}\right)^{-4} \left(\frac{T_{\mathrm{eff}}}{T_{\mathrm{eff,}\odot}}\right)^{3/2}\\
\frac{M}{\mathrm{M}_\odot} & = & \left(\frac{\Delta \nu}{f_{\Delta \nu}\Delta \nu _{\odot}}\right)^{2} \left(\frac{L}{L_{\odot}}\right)^{3/2} \left(\frac{T_{\mathrm{eff}}}{T_{\mathrm{eff,}\odot}}\right)^{-6}\\
\frac{M}{\mathrm{M}_\odot} & = & \left(\frac{\nu _{\mathrm{max}}}{f_{\nu _{\mathrm{max}}}\nu _{\mathrm{max,}\odot}}\right) \left(\frac{L}{L_{\odot}}\right) \left(\frac{T_{\mathrm{eff}}}{T_{\mathrm{eff,}\odot}}\right)^{-7/2}\\
\frac{M}{\mathrm{M}_\odot} & = & \left(\frac{\nu _{\mathrm{max}}}{f_{\nu _{\mathrm{max}}}\nu _{\mathrm{max,}\odot}}\right)^{12/5} \left(\frac{\Delta \nu}{f_{\Delta \nu}\Delta \nu _{\odot}}\right)^{-14/5} \left(\frac{L}{L_{\odot}}\right)^{3/10}
\end{eqnarray}

These mass equations are not independent, since they are just different combinations of Eqns. (1), (2) and the Stefan-Boltzmann law. Therefore, if there were no measurement errors all equations would give the same mass, and the scaling relation radius would be equal to that from the Stefan-Boltzmann law. Thus, while there are four different equations for the mass, there are only two equations for the radius, which is Eqn. (1) and the Stefan-Boltzmann law, respectively. 
We gathered the mass and radius estimates in Table~\ref{tab:seispropdata} along with other estimates of mass and age from the literature that we discuss later.

Uncertainties on masses and radii of the above relations (including the mean values) were estimated by Monte Carlo simulation. Here the observables ($L$, $T_{\rm eff}$, $\Delta\nu$, and $\nu_{\rm max}$) were assumed to be normally distributed. The 1$\sigma$ uncertainties in Table~\ref{tab:data}~and~\ref{tab:seisdata} were used along with $\pm$100 K for $T_{\rm eff}$.

We have also made best fit mass and radius estimates of all available data simultaneously with a Markov chain Monte Carlo (MCMC) sampling method, using \emph{emcee} \citep{Foreman-Mackey2013}. This method makes use of three equations representing the independent observations: The Stefan-Boltzmann equation for the luminosity and the two independent asteroseismic scaling relations for $\rho$ and $\mathrm{g}$ \citep[e.g.][and references therein]{Brogaard2022}:
\begin{eqnarray}
    \frac{ \Delta\nu }{ \Delta\nu_{\odot} } & = & f_{ \Delta \nu }\left(\frac{ \rho }{ \rho_{\odot} }\right) \\
    \frac{ \nu _{\mathrm{max}} }{ \nu _{\mathrm{max}}, \odot}  & = & f_{\nu _{\mathrm{max}}} \frac{ \mathrm{g} }{ \mathrm{g}_{\odot} } \left(\frac{ T_{\mathrm{eff}} }{ T_{\mathrm{eff}, \odot} }\right)^{1/2}
\end{eqnarray}
This method assumes that no inherent bias is present in either the asteroseismic or astrometric/photometric observations, and explores the parameters assuming that the uncertainties reflect the true measurement uncertainties. We used a simple $\chi^2$ evaluation for each of the equations as a likelihood estimator. In the results reported in Table.~\ref{tab:seispropdata}, we have also ignored that Eqn. (7) and the Stefan-Boltzmann equation correlate through the effective temperature.

For KIC\,8461659, it was explored if this last assumption had a significant effect on the result, by sampling $T_{\mathrm{eff}}$ with a prior as well. The uncertainties decreased when including $T_{\rm eff}$ in the sampling, by $31\%$ for mass and $28\%$ for radius if using the sampled temperature for the Stefan-Boltzmann temperature and the scaling relation temperature for $\nu_{\rm max}$, and by $14\%$ for mass and $11\%$ for radius if \emph{also} allowing the sampled temperature to vary the bolometric correction $\rm BC_{\it Ks}$ and extinction $\rm A_{\it Ks}$. When varying both, the sampled temperature only varied by $\pm 3$K, which is unrealistic. We elected to use the results from the simpler MCMC sampling using only mass and radius, since it is likely that systematic uncertainties in the asteroseismic scaling relations, and in the bolometric correction / extinction coefficient calculation, play a more dominant role in the final uncertainties than reflected when also sampling the temperature.

For the MCMC, the results were generally very close to the results from the asteroseismic scaling relations (Eqns. 1 and 2), with only small changes induced by including the luminosity. This naturally arises from the quoted uncertainties: The asteroseismic scaling relation uncertainties are simply much smaller than the luminosity uncertainty. As can be seen when using the four different mass equations (2-5, results in Table~\ref{tab:seispropdata} and figures in Appendix B), the observed bias between the Stefan-Boltzmann equation and the scaling relations can be quite large and even change the mass/radius correlation dramatically. This means that the MCMC result does not necessarily reflect the true mass and radius of the stars, if it is based on biased measurements. And since we do not know whether the bias might stem from the luminosity or the scaling relations, it is difficult to account for.

The mean mass of the four equations is likely less sensitive to systematics than the MCMC because a potential biased asteroseismic parameter or $T_{\rm eff}$ measurement will have a smaller effect on the mean mass than on any of the individual equations. We also experimented with a weighted average of the mass and radius equations, using the Monte Carlo simulation 1D uncertainties. The weighted average generally falls between MCMC and flat mean, but usually shifted towards the MCMC results.
We comment further on the uncertainties when we compare our results to isochrones in the next section.

\section{Stellar models and isochrone comparisons}
\label{sec:comparisons}

In this Section, we carry out a detailed comparison of the derived properties of the NGC\,6866 giants to stellar evolution models. We first consider models without rotation before extending to models including rotation.

\subsection{Models without rotation}

\begin{figure}[H]
   \centering
    \includegraphics[width=\hsize]{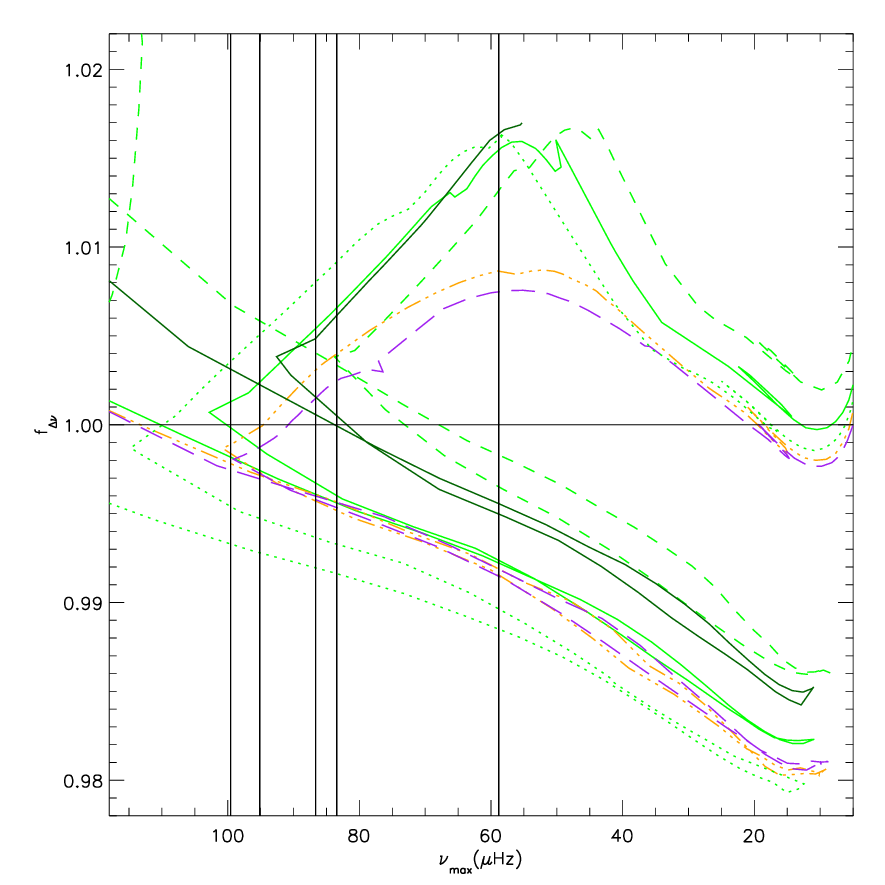}
      \caption{Predicted corrections to $\Delta\nu$ as a function of $\nu_{\rm max}$ for isochrones with details in the text and in legend of Fig.~\ref{fig:CMD} and Fig.~\ref{fig:mrt}. For clarity, not all isochrones are shown. For reference, the stable HeCB phase begins approximately at the point (93,1.004) and continues to the right for the dark green solid 0.43 Gyr isochrone. The vertical black lines mark the measured $\nu_{\rm max}$ values for the NGC\,6866 giants.}              
         \label{fig:ddeltanu}
   \end{figure}

Figs.~\ref{fig:CMD}, ~\ref{fig:ddeltanu}, ~\ref{fig:mrt} and ~\ref{fig:mr_s} shows comparisons on several planes of our and literature measurements to selected isochrones. 

Fig.~\ref{fig:CMD} shows a \textit{Gaia} CMD with stars selected to be members of NGC\,6866 by proper motions and parallaxes. Each star has been shifted in magnitude according to the individual \textit{Gaia} DR3 parallax and a mean reddening and absorption in the $G$-band corresponding to a mean reddening of $E(G_{\rm BP}-G_{\rm RP})$ = 0.19. This number was adopted based on the mean $E(B-V)$ of the giants and the 25 brightest main-sequence stars using the \citet{Green2019} reddening map. 
All stars had parallaxes corrected using the mean correction value for the six giants as calculated using the \citet{Lindegren2021} procedure and software. The red giants are shown both using this procedure and also using their individual parallaxes, corrections to the parallax, and $E(B-V)$ values. Using this latter procedure, the individual giants are marked with red symbols of different kind also mentioned at the top of Table~\ref{tab:data} and used in later figures for easy cross-reference. As seen, the giants, which are all expected to be HeCB stars according to asteroseismology, align in colour much closer to expectations from the isochrones when individual reddening values are used, suggesting a significant amount of differential reddening in the direction of NGC\,6866. The isochrones shown are from the PARSEC v1.2S grid \citep{Bressan2012}, the PARSEC v2.0 grid \citep{Nguyen2022} and MESA \citep{Paxton2011,Paxton2013} model grids of \citet{Rodrigues2017,Miglio2021,Campante2017,North2017} and our own extensions thereof, for which we calculated isochrones using the software of \citet{Dotter2016} and the \textit{Gaia} magnitudes using the YBC website\footnote{http://stev.oapd.inaf.it/YBC/index.html} \citep{Chen2019}. The details of the MESA models are given in the respective papers. Since we will investigate the effect on the models of convective-core overshoot during the main sequence, we repeat here that for the MESA models, overshooting was applied during the main sequence in accordance with the \citet{Maeder1975} step function scheme. The overshooting parameter is $\alpha_{\rm ovH} = \beta \cdot H_p$ where $H_p$ is the pressure scale height and $\beta = 0.2$ is the default value. The specific composition, age and core overshoot parameter for each isochrone is given in the legend of Fig.~\ref{fig:mrt}. 

In the upper panel of  Fig.~\ref{fig:mrt} we show the asteroseismic masses and radii of the giants compared to the same isochrones as in Fig.~\ref{fig:CMD}. The black filled circles mark the best estimates of the measured values from Eqns. (1) and (2) with $f_{\nu_{\rm max}}$ = 1 and $f_{\Delta \nu}$ estimated from Fig.~\ref{fig:ddeltanu} and given in Table~\ref{tab:data}. Each giant is also marked with the same symbol as in the other figures and as mentioned in Table~\ref{tab:data}. For each star, three positions are marked in red in addition to the best estimate, corresponding to adding, in turn, 1$\sigma$ to each of the parameters $T_{\rm eff}$, $\Delta \nu$, and $\nu_{\rm max}$ using 100K for $T_{\rm eff}$ uncertainty and the numbers in Table~\ref{tab:data} for the asteroseismic uncertainties. This means that from left to right for a given target, the symbol marks mass and radius calculated with an additional 1$\sigma$ having been added to first $\Delta \nu$, then to no parameter (best estimate), then to $\nu_{\rm max}$ and finally to $T_{\rm eff}$. 
Each star is also represented by a corresponding blue symbol marking its mean mass from equations (2)-(5) and the radius from the Stefan-Boltzmann law with our derived luminosity and the spectroscopic $T_{\rm eff}$ values from APOGEE DR17.
Four of the five giant stars have very similar radii, and since all the isochrones suggest a linear mass-radius relation at the beginning of the HeCB phase, we also mark with plus signs the mean of the masses and radii of these four stars. The red plus sign is the mean of their parameters from Eqn.~\ref{eq:01} and Eqn.~\ref{eq:02}. The blue plus sign is the mean of the mean masses from the four mass equations for the four stars, while the radius is the mean of the radii from the Stefan-Boltzmann equation for the same stars. The blue and red plus signs differ by less than 0.04 $M_{\odot}$ and 0.1 $R_{\odot}$, signalling very good mass and radius consistency in the mean, although the measurements of each individual star is less precise. The difference in scatter of the red and blue points for the individual stars demonstrates that using the luminosity constraint decreases the random uncertainty for the individual stars; comparing to an imaginary isochrone with a slope similar to those shown, but matching all stars as well as possible, suggests a random mass uncertainty of more than 0.1 $M_{\odot}$ for each individual star without the luminosity constraint (black points), but less than 0.05 $M_{\odot}$ for each individual star with it (blue points). For our best age estimate later, we will therefore require the isochrone to match the mass-radius point of the blue plus sign within $\pm0.05 M_{\odot}$. This mass precision (1.8\%) is significantly better than what is usually obtained for single giant stars \citep{Li2022, Campante2023}.   

Several isochrones are shown in the mass-radius diagram in the upper panel of Fig.~\ref{fig:mrt} for comparisons to the measurements. Starting at the lower left they show the upper main-sequence, shifting to the subgiant and red giant phases where the isochrones become almost vertical. The isochrones then continue out of the figure at the top with the upper RGB phase, and returns with the lower part of the fast-descending beginning of the core-helium burning phase. This phase ends in the stable HeCB phase, which is longer-lived and therefore the isochrones only bends upwards slowly on this latter part in the mass-radius diagram. Since we know from the asteroseismic analysis that the measured stars are in the HeCB phase of evolution it is only this latter part of the isochrone that is expected to match the observations. 

We observe from the mass-radius diagram in the top panel of Fig.~\ref{fig:mr_s} that the solid blue PARSEC v1.2 isochrone \citep{Bressan2012} significantly overpredicts the minimum radius in the HeCB phase for NGC\,6866. This overprediction is significantly reduced with the PARSEC v2.0 isochrone (blue dash-dotted), which has a smaller core overshoot parameter than in v1.2 and also a changed mixing scheme for convection zones from 'instantaneous' to diffusive \citep[see section 3.4 of][]{Nguyen2022}. While the PARSEC v2.0 value for the core overshoot parameter was calibrated using eclipsing binary stars close to the end of the main sequence \citep{Costa2019}, it is still too large to allow the isochrones to match our asteroseismic measurements of the NGC\,6866 HeCB stars. This is implying that the He core in the HeCB phase is too large in these models and that one way to reduce its size is by reducing the amount of core overshooting. We therefore continue our investigation here with other isochrones, where we can control the amount of core overshooting, while looking into rotation in Sect.~\ref{sec:rot}.

The green dashed isochrone in the top panel of Fig.~\ref{fig:mr_s} is from the MESA stellar models generated and used by \citet{Rodrigues2017} for an age of 0.50 Gyr and solar metallicity. This isochrone does a better job of matching the measured masses and radii than the PARSEC isochrones, and thus appears to be an improved representation of the cluster stars. However, in Fig.~\ref{fig:ddeltanu} one can see that this isochrone does not reach the stars with the highest observed $\nu _{\mathrm{max}}$ values for the HeCB phase, and thus cannot be a true representation of these. This isochrone assumes a convective-core overshoot parameter of $\beta = $ 0.2. With additional isochrones from the same model grid, differing by having a higher age or a lower core-convective overshoot parameter, Fig.~\ref{fig:ddeltanu} and panels 2 and 3 of Fig.~\ref{fig:mr_s} show that only a lower overshoot isochrone achieves a solution consistent with all the measurements. On the mass-$T_{\rm eff}$ plane in Fig. ~\ref{fig:mrt} the isochrones remain slightly cooler than our measured values. This is true for all but the gold-coloured isochrone, which is similar to the dark green full-drawn isochrone in all aspects except that it includes atomic diffusion (and thus the solar calibration results in a larger mixing length parameter, and different values of X, Y, and Z, see details in \citealt{Miglio2021}). Specifically, the isochrone has the same assumed Z=0.01756 as the solar metallicity cases without diffusion, but a larger $Y=0.271$ and mixing length parameter $\alpha_{\rm  MLT} = 2.12$ compared to $Y=0.266$ and $\alpha_{\rm  MLT} = 1.96$ for the isochrones without diffusion \footnote{Actually, the models including diffusion were calculated with diffusion turned off, but with the mixing-length and helium content calibrated on a solar model including diffusion. We checked for a $2.3 M_{\odot}$ star that the evolutionary track differences are insignificant. Thus, the major effects of including diffusion in the models are the indirect effects caused by the differences in the solar calibration.}. As seen in the fourth panel of Fig.~\ref{fig:mr_s}, the mass-radius relation is unaffected by assumptions on diffusion.

In the CMD in Fig.~\ref{fig:CMD}, the comparison of the HeCB stars to the isochrones also support the conclusion that only a low value of the overshoot parameter can reach a low enough luminosity at the beginning of the HeCB phase. This is the same as a small enough radius at the measured mass, given that the $T_{\rm eff}$ remains at the same value throughout most of the HeCB phase. The isochrone including diffusion seems to match the colour of the HeCB phase worse than the non-diffusion isochrone in Fig.~\ref{fig:CMD}. However, such a conclusion depends on the exact [Fe/H] and reddening values well within their uncertainties as well as on uncertainties in colour-$T_{\rm eff}$ relations. The latter are different between the isochrones and the derivations of the photometric $T_{\rm eff}$ values because the \citet{Casagrande2018} relations are based on the MARCS \citep{Gustafsson2008} atmosphere grid that does not extend hotter than 8000 K and thus cannot be used to reproduce the upper main sequence.  

\begin{figure}[H]
   \centering
    \includegraphics[width=\hsize]{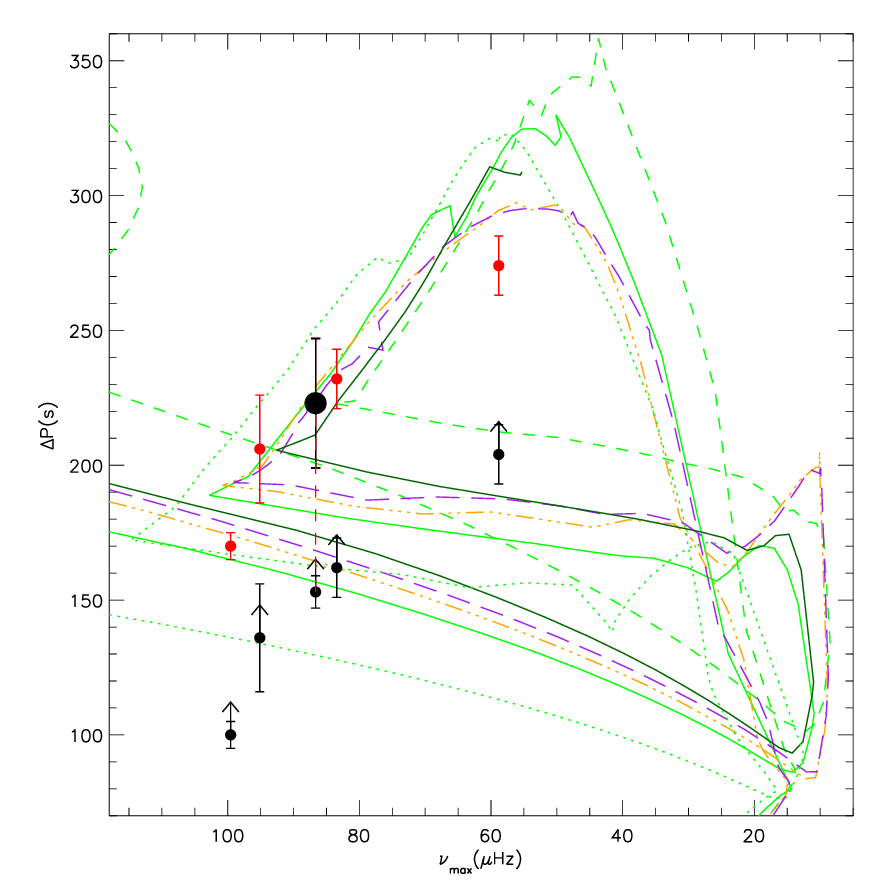}
      \caption{Predicted and observed period spacings as a function of $\nu_{\rm max}$. Isochrone details are in the text and in legend of Fig.~\ref{fig:CMD} and Fig.~\ref{fig:mrt}. For clarity, not all isochrones are shown. For reference, the isochrone RGB phase begins on the left-hand side and reaches the RGB tip on the lower right before returning left through the contracting HeCB phase, then taking another shift to the right at the beginning of the stable HeCB phase. This happens at the point (93,208) for the dark green full-drawn thick 0.43 Gyr isochrone. The black arrows mark the observed period spacings which are extreme lower limits of their asymptotic counterparts. The red dashed line connects the observed and asymptotic period spacing for the one star that has both measurements. The red circles are the observed period spacings shifted by the amount of the red dashed line, and represent an approximation of the asymptotic period spacings assuming that the difference between the observed and asymptotic period spacing is similar for all five stars.}              
         \label{fig:deltaP}
   \end{figure}

\begin{figure}[H]
   \centering
    \includegraphics[width=\hsize]{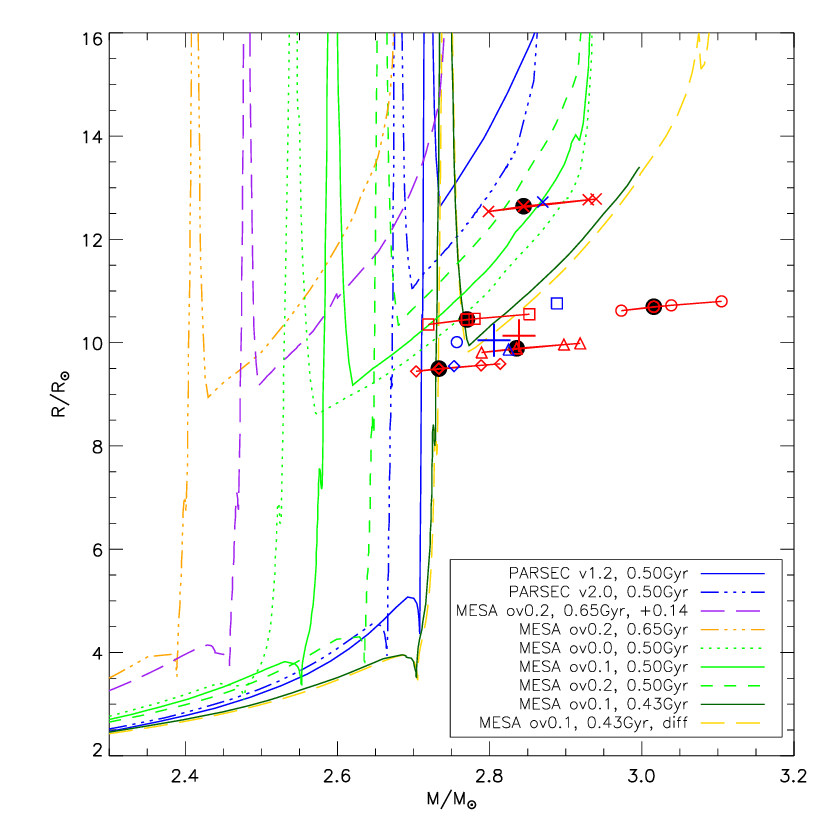}
    \includegraphics[width=\hsize]{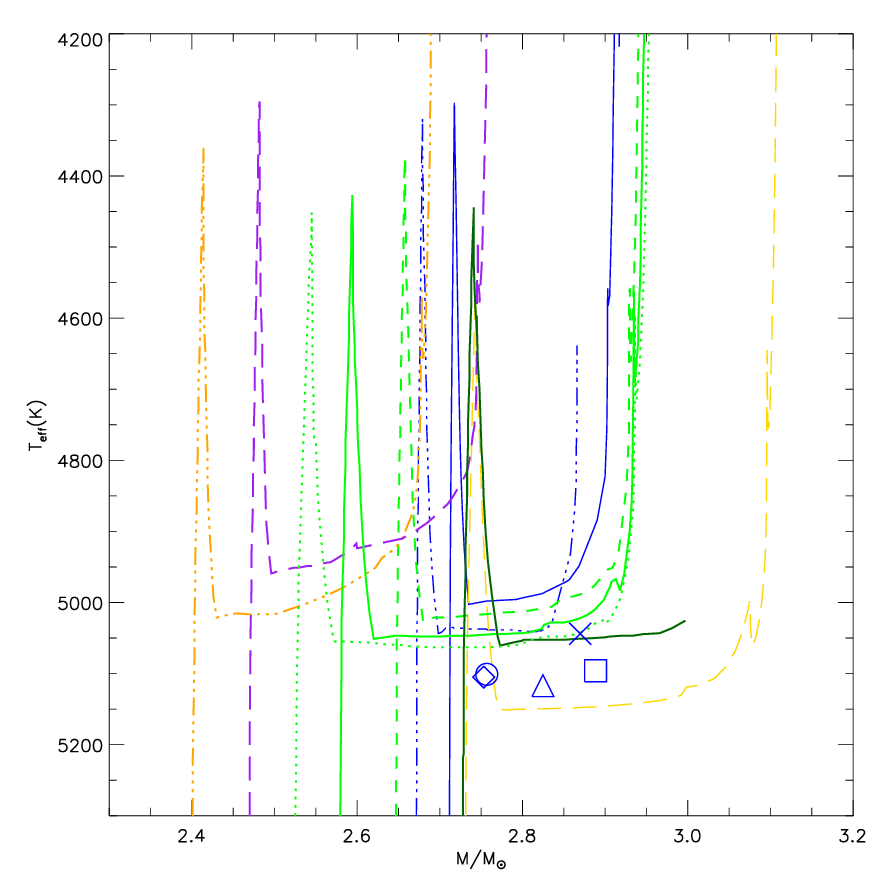}
      \caption{Mass-radius and mass-$T_{\rm eff}$ diagrams of the NGC\,6866 giants.\\
      \textit{Upper panel}: Mass-radius diagram. Black filled circles mark the measured masses and radii of the NGC\,6866 giants according to Eqns. (1) and (2). Red symbols mark the same values and those corresponding to adding, in turn, 1$\sigma$ to $\Delta\nu$ (leftmost point), $\nu_{\rm max}$, and $T_{\rm eff}$ (rightmost point). Blue symbols mark the mean masses of Eqns. (2)-(5) and radii from the Stefan-Boltzmann equation. The red and blue plus signs mark the mean masses and radii of the red and blue symbols, respectively, not counting the most evolved star. Details of the isochrones are given in the legend. 
      \\ \textit{Lower panel}: Mass-$T_{\rm eff}$ diagram. Isochrones are the same as in the upper panel. Blue symbols mark the mean masses of Eqns. (2)-(5) and $T_{\rm eff}$ values from APOGEE DR17.}
         \label{fig:mrt}
   \end{figure}

\begin{figure}[H]
   \centering
    \includegraphics[width=\hsize]{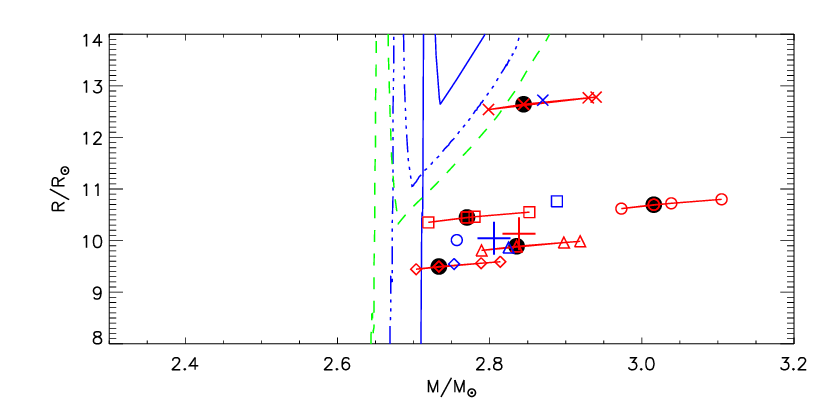}
    \includegraphics[width=\hsize]{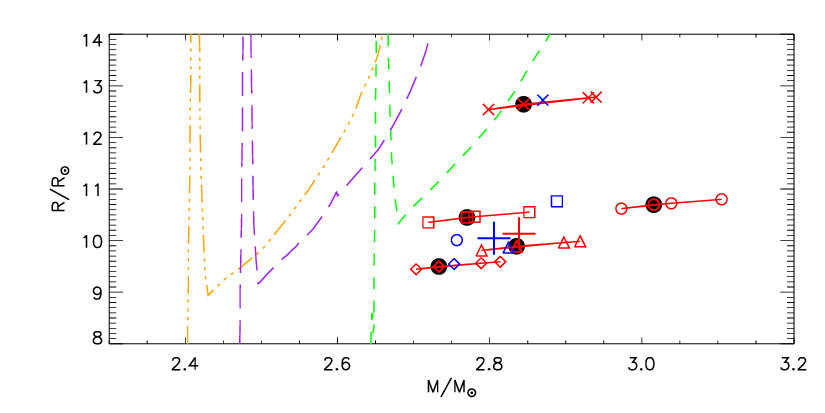}
    \includegraphics[width=\hsize]{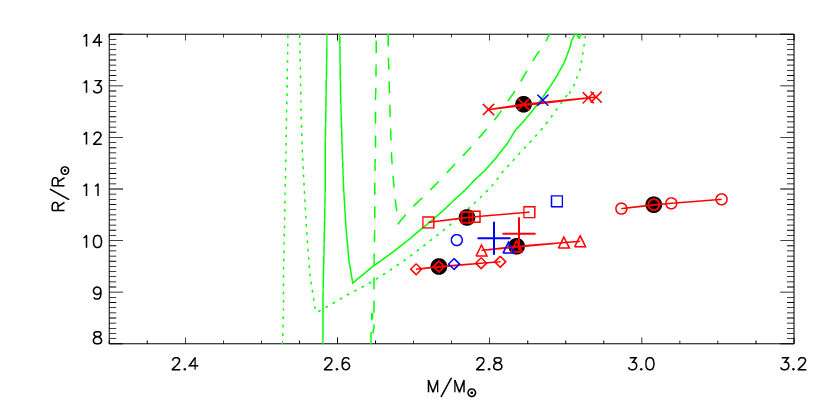}
    \includegraphics[width=\hsize]{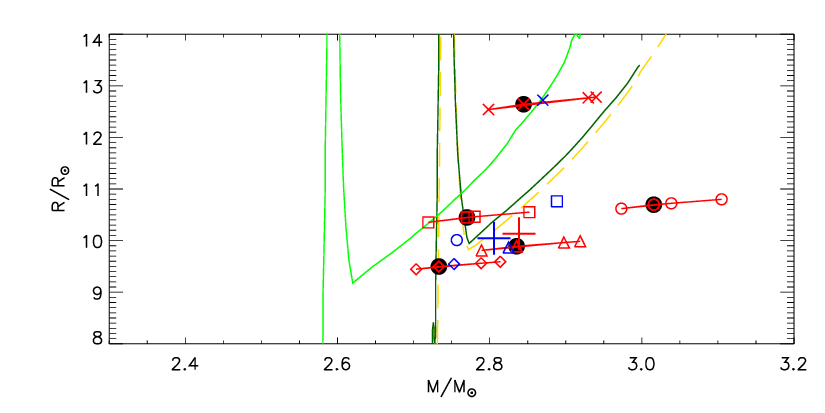}
    \includegraphics[width=\hsize]{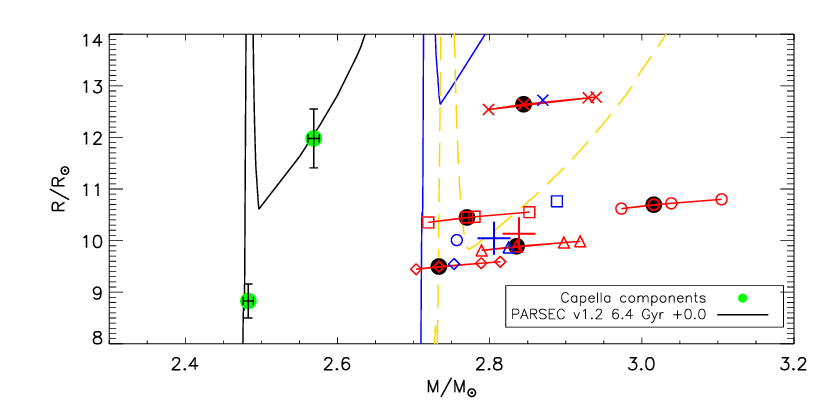}            
      \caption{Mass-radius diagrams. As Fig.~\ref{fig:mrt} but showing different subsets of the isochrones. Details of the isochrones are given in the legend of Fig.~\ref{fig:mrt}. From top to bottom, the panels show: (a) Effects of different isochrone sets, (b) effects of age and metallicity, (c) effects of changing core-overshoot efficiency, (d) best age estimates and negligible effects of diffusion in the mass-radius relation, and (e) comparison to parameters of Capella. 
      \textit{Bottom panel}: Green circles mark the masses and radii of Capella measured by \citet{Torres2015}, see Sect.~\ref{sec:comparisons}. The lower mass component is in the Hertzsprung gap and the higher mass component is in the HeCB phase.
      }
         \label{fig:mr_s}
   \end{figure}

\begin{figure}[H]
   \centering
    \includegraphics[width=\hsize]{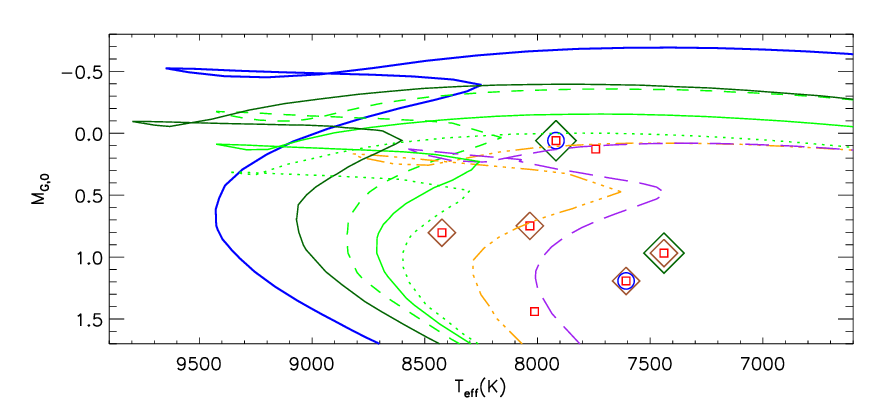}
      \caption{Hertzsprung-Russel diagram zoom on the upper main-sequence. As in the lower left panel of Fig.~\ref{fig:CMD} but with the x-axis colour replaced by $T_{\rm eff}$, and only showing stars with spectroscopic $T_{\rm eff}$ measurements in \citet{Frasca2016}. For clarity, only some of the isochrones from Fig.~\ref{fig:CMD} are repeated.}
         \label{fig:to_teff}
   \end{figure}

Attempts to discriminate between the isochrones based on the turn-off region of the CMD are complicated by difficulties in identifying how to match an isochrone to sparse and scattered observations. Of the 25 brightest stars at the turn-off in the CMD, marked with small red squares in Fig.~\ref{fig:CMD}, only five are not listed as either $\delta$\,Sct, $\gamma$ Dor, or rotational variables on SIMBAD. Among these 25 turn-off stars, most have \textit{vbroad} values in \textit{Gaia} DR3, and they are in the range 37-260 km/s, but with huge uncertainties of similar size as the values themselves, or sometimes even larger. Five stars, marked with blue circles in the lower-left panel of Fig.~\ref{fig:CMD} are identified as \textit{photometric variable}. Four stars, marked with brown diamonds, have measured $v \sin i$ values, which are all close to 150 km/s \citep{Frasca2016}, while three, marked with light green diamonds, were measured as photometric rotators with rotation periods of 2-5 days by \citet{Nielsen2013}. 
Unlike the situation for the giant stars, attempts to use individual reddening values and/or parallaxes for the turn-off stars did not tighten the cluster CMD sequence, suggesting that differential reddening is not the main cause for the apparent photometric scatter. This is supported by a comparison of the lower left panel of Fig.~\ref{fig:CMD} to Fig.~\ref{fig:to_teff} where we replaced the x-axis colour by $T_{\rm eff}$ for stars with a spectroscopic temperature measurement in \citet{Frasca2016}; roughly the same pattern appears in colour and $T_{\rm eff}$, showing that the main colour-differences are intrinsic to the stars, and not caused by differential reddening or photometric blending due to binarity of some of the stars (at least not those with spectroscopic $T_{\rm eff}$ measurements). Instead, the photometric scatter could potentially be explained by the relatively fast rotation of the turn-off stars in combination with different orientations of their inclination angles. However, the four stars with very similar $v$sin$i$ measurements span a broad range in both colour and $T_{\rm eff}$, which disfavours such an explanation. 
The brightest star on the hot side of the Hertzsprung gap, marked with a purple circle in Fig.~\ref{fig:CMD} is a known $\delta$\,Sct and binary star system \citep{Murphy2018} that should likely not be matched by the isochrones, and there could be other unknown binaries among the brightest stars. Time-scale arguments suggest that the vast majority of these stars should be matched by the part of the isochrone that ends at the reddest (coolest) and brightest point on the upper main-sequence before it bends back to the blue (hotter) during the fast contraction phase. However, we cannot be sure whether a few stars actually crossed that point. As seen in Figs.~\ref{fig:CMD} and ~\ref{fig:to_teff}, this seems to be the case for our MESA isochrones with an age close to 0.43 Gyr and an core-overshoot parameter of $\beta=0.1$. The few that are potentially problematic are also photometrically variable stars that need further investigation before they can be used to disfavour an isochrone. Values of $\beta$ much below 0.1 cause increasing tension, with an increasing number of stars becoming brighter than the isochrone luminosity at the end of the main-sequence.

A comparison of the lower left panel of Fig.~\ref{fig:CMD} to Fig.~\ref{fig:to_teff} demonstrates that the isochrones are bluer than the observations because they are also hotter than the observations, and the colour mismatch is thus not related to potential issues with colour-temperature relations. It can also be inferred that this could be alleviated by increasing the isochrone metallicity, although not by enough to remove the $T_{\rm eff}$ tension while also respecting the spectroscopic metallicity measurements within uncertainties. We comment on potential causes for $T_{\rm eff}$ issues later. 

A more complete and detailed observational study that measures rotation speeds, spectroscopic $T_{\rm eff}$, and binarity in the turn-off region of NGC\,6866 could potentially help clarify issues at the CMD turn-off. 

In conclusion, the MESA isochrone that best matches all observations is one that has a convective-core overshoot parameter of about $\beta = $ 0.10 or less, although all isochrones are too blue/hot in the turn-off region of the CMD in Fig.~\ref{fig:CMD}. We show in our figures an example for solar metallicity, 0.43 Gyr and $\beta = $ 0.10 both with and without diffusion. As seen in Fig.~\ref{fig:ddeltanu}, this particular isochrone without diffusion does not quite reach high-enough values of $\nu_{\rm max}$. Increasing the metallicity and age, including diffusion, and reducing the core-overshoot parameter would all help to reduce that problem. Thus solutions exist with solar or higher metallicity, and/or including diffusion, with $\beta$ appropriately adjusted from 0 to slightly below 0.1. But no good solutions exist for $\beta > 0.1$ for any realistic choices of the metallicity. Values of $\beta$ much below 0.1 seem disfavoured by the brightest stars on the main sequence.  

Comparing the mass difference in the HeCB phase between isochrones of different age or composition to the value and uncertainty of the asteroseismically measured mean HeCB mass in the mass-radius diagram in Fig.~\ref{fig:mrt} yields a best estimate age of $0.43\pm0.03$ Gyr for a fixed solar metallicity, increasing by an additional $\pm0.02$ when also allowing for a $\pm0.1$ dex uncertainty on metallicity for NGC\,6866.

We find a preference for a convective-core overshoot parameter significantly lower than the $\beta = $ 0.2, which is often used as a standard value in stellar models. A value slightly lower than the standard one, but not as small as suggested by our current study is supported by the analysis of multiple eclipsing binary stars in the older open cluster NGC\,2506 by \citet{Knudstrup2020}. In general, calibrations of core-overshoot using eclipsing and non-eclipsing binary field stars support a higher constant value of $\beta = $ 0.2 for stars with masses above 2 $M_{\odot}$ \citep{Costa2019, Claret2018} including the mass range of the NGC\,6866 giants. With the exception of Capella \citep{Torres2015}, all those calibrators \citep{Claret2017} with masses similar to the NGC\,6866 giants are unfortunately much bigger, potentially even more evolved, and many without a metallicity measurement, likely because they are located in the SMC. Thus, there are further complications when trying to use those as calibrators of core-overshooting on the main sequence. As just one example, \citet{Bossini2017} showed that the core-overshooting efficiency in the HeCB phase needs to be significantly larger than $\beta = $ 0.2 to match asteroseismic observations in the open clusters NGC\,6819 and NGC\,6791, and thus different from the core-overshooting during the main sequence. Such a complication will affect the use of any giant in an evolved HeCB phase or later, for core-overshoot calibration on the main-sequence, including Capella \citep{Marini2023}, given that its properties reveal its primary component to be in the HeCB stage. If such issues have affected previous attempts to calibrate core-overshooting efficiency on the main-sequence, then perhaps the main-sequence core-overshoot efficiency is generally lower at the mass of the NGC\,6866 giants than at lower masses. However, a comparison between the NGC\,6866 giants and the components of Capella in the bottom panel of Fig.~\ref{fig:mr_s} shows that our models cannot match both cases; for Capella, the main constraint on core overshoot comes from the relatively small mass difference between the lower mass Hertzsprung gap component and the higher mass HeCB component, and thus core overshoot needs to be increased to move upwards the beginning of the stable HeCB phase where the HeCB stars have their minimum size. This is inconsistent with the constraint on the minimum size of the HeCB stars in NGC6866. 

A potential, but purely speculative solution to the apparent discrepancy could be related to interactions between core-overshoot and rotation (cf. Sect~\ref{sec:rot}) with the companion influencing the latter in the case of binary stars. While this remains speculative, it could potentially remove the apparent discrepancy between the NGC\,6866 giants and Capella (and other binary stars).

There are other potential issues with stellar models that might influence both our results and those of others. Among these are changes to the mixing length parameter and T-$\tau$ relations. It is known from 3D simulations of stellar atmospheres that these depend on stellar properties \citep{Trampedach2014a,Trampedach2014b,Magic2015}, but this is still not generally accounted for in stellar models, partly because the simulations do not yet cover all relevant ranges of stellar parameters. \citet{Mosumgaard2018,Mosumgaard2020} did make examples of such models for a limited mass-range at solar metallicity. Their evolutionary track of a 1.4 $M_{\odot}$ model, the largest mass they consider, is offset to slightly cooler $T_{\rm eff}$ during the main sequence and slightly hotter $T_{\rm eff}$ of the giant phases. These differences remain well below 100 K during all phases of evolution and the time on the main sequence is largely unaffected. Unfortunately, we cannot know if this picture remains the same for the higher masses representing NGC\,6866.

An exhaustive investigation of convective-core overshooting remains outside the scope of the present paper simply because we do not have neither the required observations nor models to settle the issue. A detailed revisit of the investigation by \citet{Knudstrup2020} and other binary stars along with a simultaneous study of asteroseismic and binary measurements in NGC\,6866, Hyades \citep{Brogaard2021a}, M44 \citep{Morales2022}, NGC\,6811 \citep{Arentoft2017}, NGC\,6819 \citep{Handberg2017}, NGC1817 \citep{Sandquist2020} and additional young open clusters could provide more constraints and insights in the future.  
\subsection{Models with rotation}
\label{sec:rot}
\begin{figure}[H]
   \centering
    \includegraphics[width=\hsize]{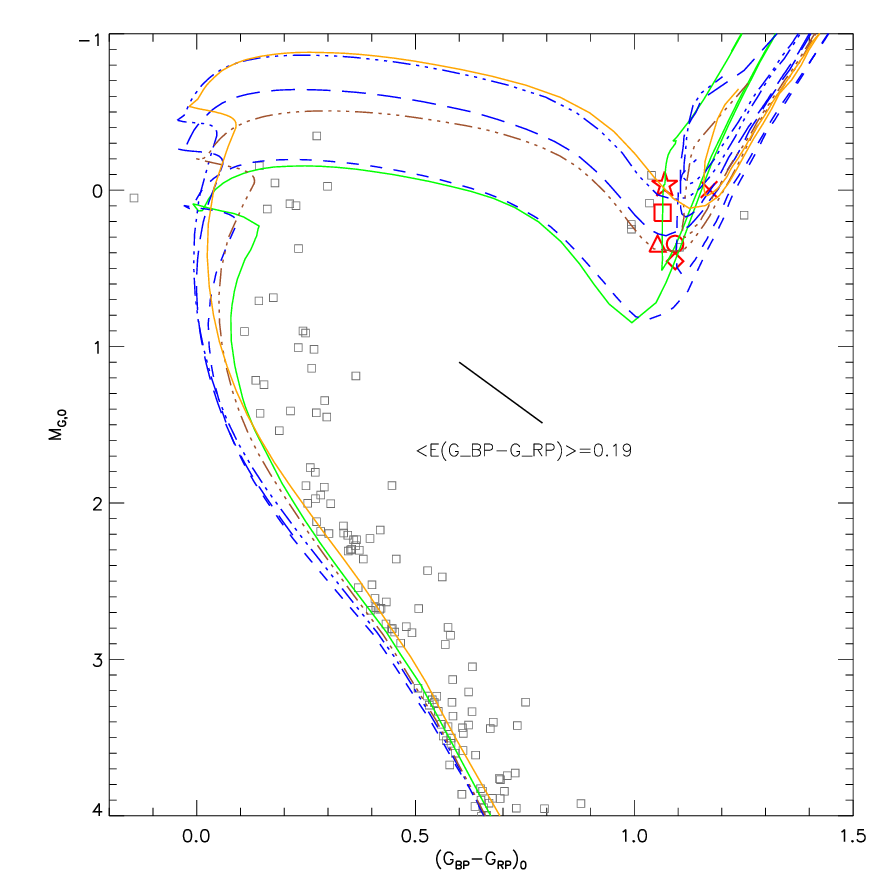}
    \includegraphics[width=\hsize]{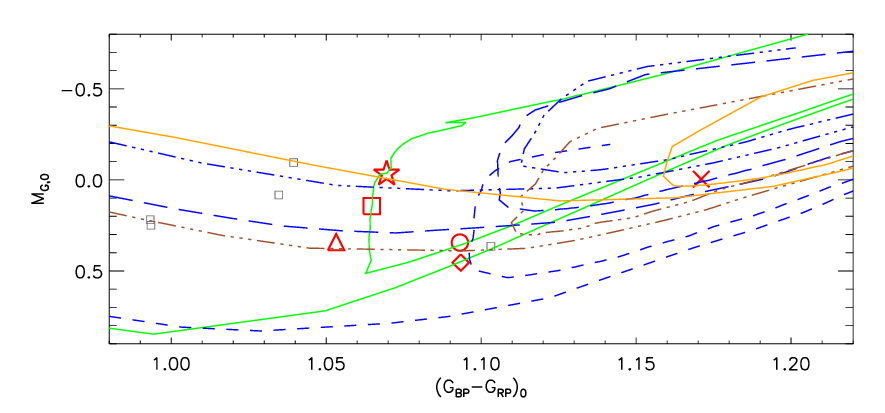}
      \caption{As Fig.~\ref{fig:CMD} but compared to Geneva isochrones including rotation. Details are given in the text and in the legend of Fig.~\ref{fig:mrt_rot}. The green full-drawn MESA isochrone is repeated for easy comparison to other figures.}
         \label{fig:cmd_rot}
   \end{figure}

The effect of rotation on 1D stellar models is known to be similar, though not identical to, that of convective-core overshooting. This is because chemical gradients left by step overshooting are different from those produced by a diffusive process such as chemical transport in shellular rotation. \citet{Eggenberger2010} investigated the differences between evolutionary tracks including either diffusion or rotation in the colour-magnitude diagram. They conclude that "the
main-sequence widening and the increase of the H-burning lifetime induced by rotation are well reproduced by non-rotating models with an overshooting parameter of 0.1, while the increase of luminosity during the post-main sequence evolution
is better reproduced by non-rotating models with overshooting parameters twice as large". Unfortunately, they do not investigate how this compares to models that include both overshoot and rotation.
Our measurements for the NGC\,6866 HeCB stars suggest an overshoot value of $\beta = $ 0.1 or lower for our MESA models. Combining this with the results of \citet{Eggenberger2010} then suggests that the true effects of rotation are smaller than predicted by 1D models. 
We attempted direct comparisons with models including rotation from \citet{Georgy2013, Ekstrom2012, Yusof2022}. Fig.~\ref{fig:cmd_rot} shows comparisons between the \textit{Gaia} CMD of the NGC\,6866 cluster members and some of these Geneva isochrones\footnote{Downloaded from https://www.unige.ch/sciences/astro/evolution/en/\\database/syclist/} for different compositions and initial rotation rates. In order to minimise differences to our MESA isochrones, we calculated the \textit{Gaia} magnitudes for the Geneva models using the YBC website \citep{Chen2019} in the same way that we did for the MESA isochrones.

\begin{figure}[H]
   \centering
    \includegraphics[width=\hsize]{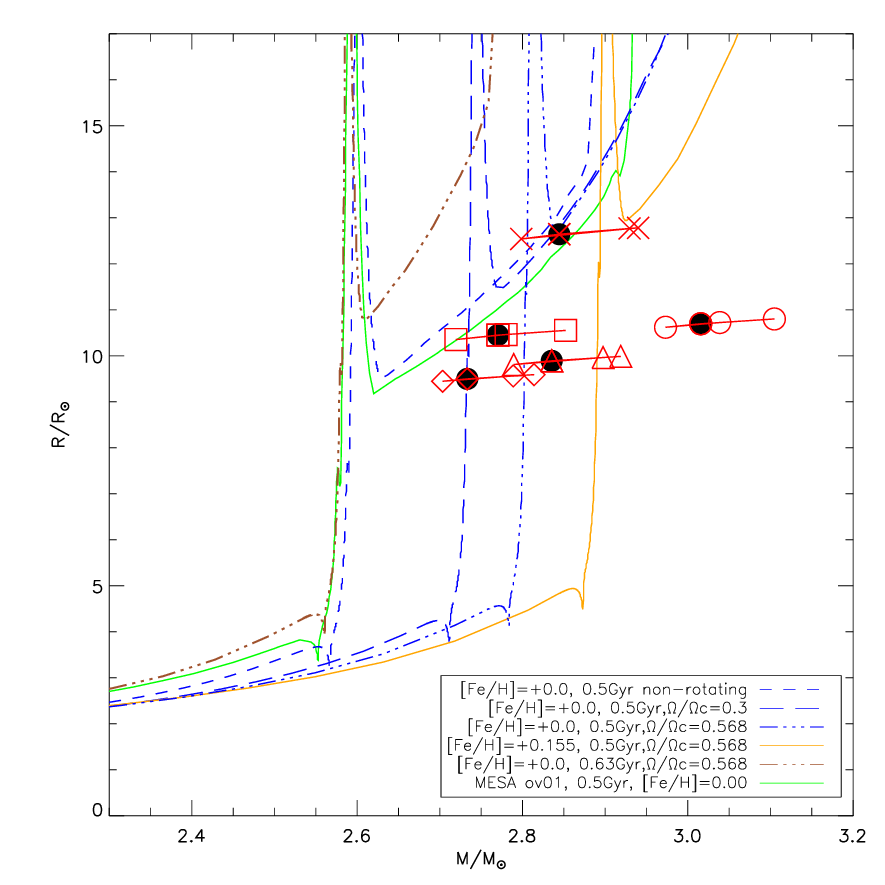}
    \includegraphics[width=\hsize]{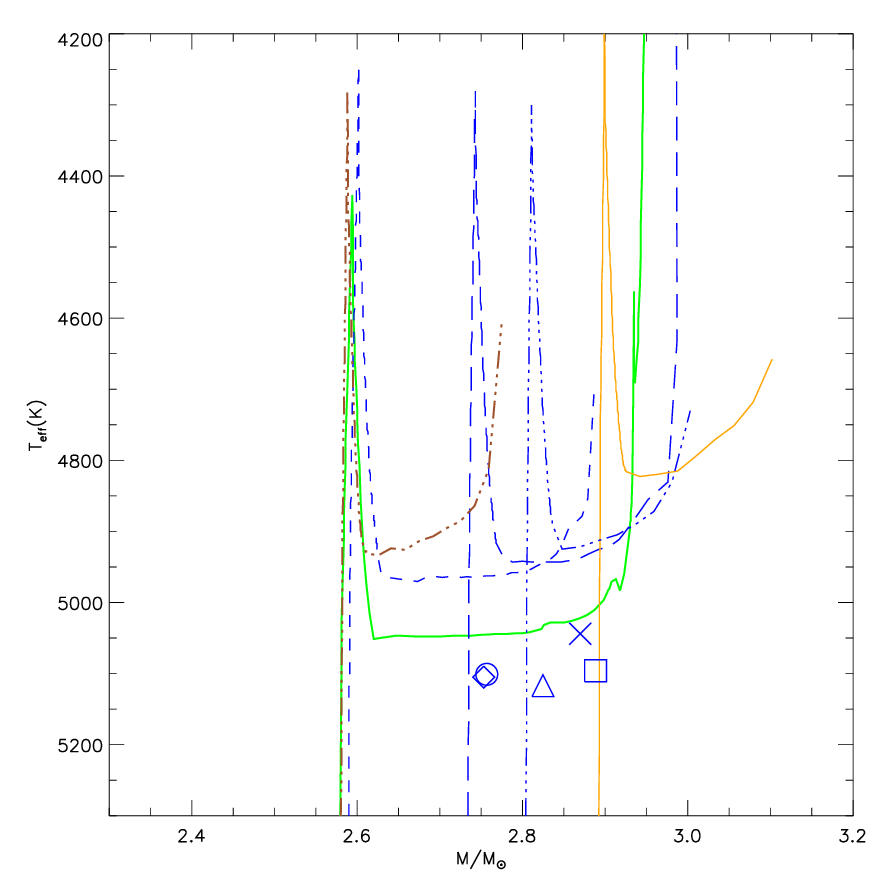}
    \caption{As Fig.~\ref{fig:mrt} but compared to Geneva isochrones including rotation. The dark green full-drawn MESA isochrone is repeated for easy comparison to other figures.}
         \label{fig:mrt_rot}
   \end{figure}

As seen in Fig.~\ref{fig:cmd_rot}, the Geneva isochrones seem too hot at the turn-off and simultaneously too cool at the HeCB phase to match the observations. The same is reflected on the mass-$T_{\rm eff}$ plane in Fig.~\ref{fig:mrt_rot}. Therefore, it would seem difficult to make firm conclusions from the mass-radius comparison in Fig.~\ref{fig:mrt_rot}. However, in Figs.~\ref{fig:cmd_rot}-\ref{fig:mrt_rot} we include also the MESA isochrone with core overshoot parameter of $\beta = $ 0.1 (and no rotation) from the previous figures for comparison. Since the Geneva isochrone without rotation matches almost exactly our MESA isochrone with core overshoot parameter of $\beta = $ 0.1 (and no rotation) from the previous figures, in this case we can make some useful conclusions, as follows. 

It can be seen in Fig.~\ref{fig:mrt_rot} that the isochrone radii for the beginning of the stable HeCB phase become too large for anything but very modest rotation, as expected from the \citet{Eggenberger2010} investigation if adopting the low value of core overshoot that we found.  It might appear from Fig.~\ref{fig:mrt_rot} that there is hardly room for any rotation. However, these models of \citet{Georgy2013, Ekstrom2012, Yusof2022} include both rotation and core overshooting with $\beta =$ 0.1. Therefore, reducing the overshoot even further might give way for some effects of rotation, although it seems limited. 

As can be seen by comparing the isochrones in the upper panel of Fig.~\ref{fig:mrt_rot}, the Geneva model radii for the HeCB phase are already slightly too large without rotation, and it only gets worse by including rotation and attempting to compensate by choosing a higher age. Since the above mentioned measured $v$sin$i$ = 150 km/s for the stars on the upper main sequence of NGC\,6866 \citep{Frasca2016} is in rough agreement with the initial surface rotation rate of the ($\Omega / \Omega_C = 0.4$) models according to fig. 2 of \citet{Yusof2022}, reducing the initial rotation rate of the models does also not seem like a possible solution. 

These results suggests that the extent of the mixed region should be equivalent to that corresponding to a moderate classical step convective overshooting, but characterising the transport processes leading to such mixing is still an open problem. 
Above referred works use shellular rotation to model the transport of angular momentum and chemicals. However, asteroseismology has revealed that such treatment of rotation alone is neither able to reproduce the solar internal rotation profile \citep[see e.g.][and references therein]{Thompson2003} nor the evolution of that profile from the MS to the red giant phase \citep[e.g.][]{Beck2012, Eggenberger2012, Cellier2013, Marques2013, Cantiello2014, Benomar2015, DiMauro2016}. 
The available ultra-precise photometry data tell us that it is no longer possible to treat phenomena linked to Convective Boundary Mixing such as overshooting, penetrative convection, entrainment, internal gravity wave generation, and rotation as independent (or additive) processes.  On the contrary, there are important interactions/feedback  among them. 

2D/3D numerical simulations of convection in rotating media have shown how rotation can affect the dynamics of the convective zone and the transition region between it and the adjacent radiative regions \citep[e.g.][]{Julien1996, Julien1997, Brummell1996, Browning2004, Brun2017} indicating a decrease of the extent of the extra-mixing layer with rotation. However, it has not been possible so far to parameterize the impact of rotation on convection and the associated transport processes over evolutionary timescales that would allow us to dynamically estimate the extent and properties of the mixed region along the evolutionary track.

Recently \citet{AugustsonMathis2019} have investigated the effect of rotation on convective penetration using the heuristic model of convection in rotating systems \citep{Stevenson1979} in combination with Zahn's linear convective penetration model \citep{Zahn1991}. They conclude that the efficiency of mixing and the extent to which it occurs above a convective region can be significantly reduced in stars with convective zones characterized by a low Rossby number (indicating high rotation). The power law relationship between the depth of convection penetration and rotation they obtained is in good agreement with previous numerical simulations. They also propose a potential implementation of their convective penetration model in stellar evolution calculations, which could be incorporated into future grids of stellar models. 

\begin{table*}
\caption{Asteroseismic properties of five NGC\,6866 giant members}  
\label{tab:seisdata}      
\centering                          
\begin{tabular}{l c c c c c }        
\hline\hline                 
KIC ID  & KIC8461659 & KIC8329894 & KIC8395903  &  KIC8264549 & KIC8264592      \\
\hline                                   
Symbol in plots & cross & square & circle & triangle  & diamond  \\
$\nu_{\rm max}(\mu$Hz) & 58.83(65) & 83.44(10) & 86.65(22) & 95.13(69)  & 99.56(67)  \\
$\Delta \nu_{ps} (\mu$Hz) & 5.223(100) & 6.755(35) & 6.800(58)& 7.147(51) & 7.663(10) \\
$\Delta \nu_{0} (\mu$Hz) & 5.140(21) & 6.684(31) & 6.734(24)& 7.314(30) & 7.631(21) \\
$\delta\nu_{02} (\mu$Hz) & 0.548$\pm0.046$ &  0.909$\pm$0.100 & 0.739$\pm$0.034 & 0.774$\pm$0.082 & 1.014$\pm$0.093 \\
$\epsilon$ & 1.032$\pm$0.041  & 1.049$\pm$0.054  & 0.991$\pm$0.041 & 1.024$\pm$0.050 &  1.087$\pm$0.033 \\
$\Delta P_{\rm obs}$(s) & 204$\pm$11 & 162$\pm$11 &  153$\pm$6 & 136$\pm$20\tablefootmark{d} & 100$\pm$5  \\
$\Delta\Pi_{1}$(s) & - & - & $223\pm24$ & - & - \\
$\Delta \nu_{c} (\mu$Hz) &  5.187$\pm$0.032 & 6.729$\pm$0.003 &  6.764$\pm$0.017 &7.293$\pm$0.053 & 7.693$\pm$0.010 \\
$\epsilon_{c}$ & 0.913$\pm$0.062 & 0.942$\pm$0.00 & 0.929$\pm$0.030 & 1.048$\pm$0.088 & 0.978$\pm$0.015 \\
$\delta\nu_{02}/\Delta \nu_{0}$ & 0.107$\pm$0.009 & 0.136$\pm$0.015 & 0.110$\pm$0.005& 0.106$\pm$0.011 & 0.133$\pm$0.012 \\
\hline
$\nu_{\rm max}(\mu$Hz) lit. & 58.26\tablefootmark{b}/58.676\tablefootmark{a} & 83.20\tablefootmark{b}/82.7\tablefootmark{c} & - & - & -\\
$\Delta \nu (\mu$Hz) lit. & 5.143\tablefootmark{b}/5.167\tablefootmark{a} & 6.660\tablefootmark{b}/6.81\tablefootmark{c}  & -  &  - & - \\
\hline
\end{tabular}
\tablefoot{  
\tablefoottext{a}{APOKASC, \citet{Pinsonneault2018}.} 
\tablefoottext{b}{\citet{Yu2018}.}
\tablefoottext{c}{\citet{Balona2013}.}
\tablefoottext{d}{Based on three modes only.}
}

\end{table*}

\begin{table*}
\caption{Asteroseismic and alternative stellar properties of NGC\,6866 giant members}  
\label{tab:seispropdata}      
\centering                          
\begin{tabular}{l | c | c c c c | c }        
\hline\hline                 
KIC ID  & KIC8461659 & KIC8329894 & KIC8395903  &  KIC8264549  & KIC8264592 &  "mean"\tablefootmark{a}    \\
\hline                                   
Symbol in plots & cross & square & circle & triangle & diamond &   \\

$f_{\Delta \nu}$ & 1.014 & 1.004 & 1.004 & 1.0  & 1.0 & -\\
$R(R_{\odot})$ seis, Eqn.(1) & 12.64(21) &  10.45(14) &  10.70(13) &  9.89(15) & 9.50(12) & 10.14  \\
$R(R_{\odot}),\rm SB$ & 12.72(56) & 10.76(47)& 10.01(43) & 9.86(57)  &  9.54(42) & 10.05 \\
$R(R_{\odot}),\rm MCMC$ & 12.66(20) &	10.48(14)&	10.64(13)&	9.90(14)&	9.51(12) & 10.13\\
\hline                                   
$M(M_{\odot})$ seis, Eqn.(2) & 2.84(14)   &   2.77(10)  &    3.02(10) &    2.84(11)    &    2.73(10) & 2.84\\
$M(M_{\odot})$ seis, Eqn.(3) & 2.90(39) &   3.03(41)   &  2.47(33)    &   2.81(51) &    2.78(37) & 2.77 \\
$M(M_{\odot})$ seis, Eqn.(4) & 2.88(23) &  2.94(24)   &    2.64(21) &   2.82(31) & 2.76(21) & 2.79\\
$M(M_{\odot})$ seis, Eqn.(5) &	2.855(89) &	2.820(51) &	2.898(46) &	2.831(94) &	2.741(58) & 2.82\\
\hline
$<M(M_{\odot})$ Eqns.> &  2.87(15)  &     2.89(15)   &    2.76(12)  &     2.82(21)   &   2.75(14) & 2.80  \\

rms$_{<M(M_{\odot})\rm Eqns.>}$ & 0.02 & 0.12   &   0.25  &   0.01 &     0.02 &  -\\

$M(M_{\odot})$ MCMC &2.86(13)&	2.793(94)&	2.977(95)&	2.840(11)&	2.741(98) & 2.84\\
\hline
$ M(M_{\odot})$ "the Cannon"\tablefootmark{b}  & 2.729 & 2.895 & 2.832 & 2.712 & 2.948& -\\
$M(M_{\odot})$ "StarHorse"\tablefootmark{c}  & $2.764\pm0$ & - & $2.35_{-0}^{+0.04}$ & $2.26_{-0.105}^{+0}$  & $2.052_{-0.163}^{+0.007}$ & -\\
 age (Gyr) "the Cannon"\tablefootmark{b}  & 0.513 & 0.455 & 0.474 & 0.580  & 0.443& -\\
age (Gyr) ML\tablefootmark{d} & 1.149(368) & 0.764(207) & 0.763(178) & 0.719(225)  & 0.757(238) & -  \\
\hline                                   
\end{tabular}
\tablefoot{  
\tablefoottext{a}{flat mean of all the stars, except KIC\,8461659, which is more evolved.}
\tablefoottext{b}{\citet{Ness2016}.}
\tablefoottext{c}{From the APOGEE+\textit{Gaia} catalogue of \citet{Queiroz2023}.}
\tablefoottext{d}{Machine Learning age from \citet{Leung2023}.}
}

\end{table*}

\section{Comparisons to other mass and cluster age estimates}
\label{sec:results}

We consider our age estimate for NGC\,6866 to be more precise and accurate than previous estimates in the literature because it does not rely on uncertainties related to how one should match the stars in the turn-off region of the CMD, and because we have been able to constrain the amount of core-overshoot in the models. Also, our investigation showed that if rotation is included in the models, then core-overshoot needs to be reduced, with the result that the inferred age will remain close to the same value. With this in mind, we now compare our cluster age and HeCB mass estimates to other measurements from the literature.

\subsection{Comparisons to other age estimates}

NGC\,6866 is listed in the machine learning based open cluster catalogue of \citet{Cantat2020} with log(age) = 8.81 corresponding to 0.646 Gyr. Since the stated catalogue uncertainty is $\sigma $(log(age)) = $\pm0.1$ yielding age = 0.646$_{-0.133}^{+0.167}$ even the lower 1$\sigma$ bound is not compatible with our estimate. Their best estimate age is likely higher than ours because they used the catalogue of \citet{Bossini2019}, who give $0.78$\,Gyr for NGC\,6866, as a main part of the input ages for training the machine learning algorithm. The work of \citet{Bossini2019} is based on PARSEC isochrones, which can be seen from our Fig.~\ref{fig:CMD} to prefer a higher age when matching the CMD of both the brightest stars on the main sequence and the HeCB stars. The same arguments can explain the difference to the NGC\,6866 log(age) = 8.861 (age = 0.726 Gyr) derived by \citet{Dias2021} who also employed the PARSEC isochrones. Therefore, while the differences to our work are understood, it illustrates well how accounting for precise cluster ages such as the one derived in this work can significantly improve catalogues of cluster ages, whether based on machine learning or not. 

The importance of such improvements can be seen for example in [C/N]-age relations that use \citet{Cantat2020} ages for their calibrations. \citet{Spoo2022} used [C/N] measurements in open clusters to derive such a relation for red clump stars (including HeCB stars with masses well above the secondary clump though they do not mention it specifically, but their calibration clusters extend to young ages where this must be the case). We have listed the [C/N] values from APOGEE DR17 for our NGC\,6866 giant members in Table~\ref{tab:data} and used their mean value <[C/N]>$_{\rm NGC6866} = -0.572$ with the [C/N]-age relation in their equation 3 to obtain log(age) = $ 8.879_{-0.343}^{+0.114}$ with the limits obtained by using the smallest and largest value of [C/N], respectively. While the uncertainty is then large enough to encompass our asteroseismic age estimate, the best estimate is close to the \citet{Cantat2020} value of log(age) = 8.81. This suggests that the age offset we see between our work and that of \citet{Cantat2020} and \citet{Bossini2019} for NGC\,6866 likely extends to other young clusters with RC/HeCB giants as well, with the result that their ages are overestimated. Through [C/N]-age relations, such bias might have been extended also to field stars.

A study of the binary system 12 Com by \citet{Lam2023} used the mass and radius of the primary component to infer the age of 12 Com and the Coma Berenices open cluster. As part of their analysis, they used a comparison of the measurements of the primary component to isochrones on the mass-radius plane to conclude that the star is not in the HeCB phase, for which their isochrones predicted a much larger radius than measured (see their fig. 7). However, comparing their measured values for 12 Com A, $M = 2.64\pm0.07 M_{\odot}$, $R = 9.12\pm0.12 R_{\odot}$, to those of our measurements of the HeCB giants in NGC\,6866 shows that the 12 Com primary star could instead be a HeCB star. The reason for the discrepancy with the isochrone in the study by \citet{Lam2023} could just be that a too large core-overshoot value was chosen for the models. From an evolutionary time scale point of view, it would also be much more likely to find the star in this phase. The effective temperature of 12 Com A was investigated by \citet{Griffin2011} who found its spectrum to closely resemble that of the star 31 Vul, which has literature $T_{\rm eff}$ values between 5060 K \citep{McWilliam1990} and 5314 K \citep{Luck2007}, with a number of measurements close to 5250 \citep[e.g.][]{Frasca2018}. These values are relatively close to those of the NGC\,6866 giants and therefore consistent with a HeCB scenario for 12 Com A. However, a Hertzsprung gap or RGB star with the radius of 12 Com A would have a similar temperature, and therefore $T_{\rm eff}$ does not allow us to favour one scenario over the other.
Further study is required to establish whether our speculations are true, but the true age of 12 Com and the Coma Berenices open cluster depends strongly on this. Future asteroseismic measurements of this star from time-series spectroscopy with the robotic telescope-network SONG \citep{Grundahl2017} could perhaps provide answers. 

\subsection{Comparisons to other mass estimates}

Four of our cluster member HeCB stars are listed in the StarHorse \citep{Queiroz2018} catalogue of \citet{Queiroz2023} based on APOGEE spectroscopic results as well as \textit{Gaia} DR3 parallaxes and photometry. We list the masses they derived in Table~\ref{tab:data} along with their $1\sigma$ uncertainties as suggested by their 16\% and 84\% quantile values. As seen, the masses are significantly lower than our estimates with one exception, and the uncertainties are suspiciously small, or even zero. We suspect that these problems arise when trying to estimate ages from automatic isochrone fitting procedures for an evolutionary stage where systematic uncertainties in models can cause edge effects, since models may not cover the entire parameter space of the observed values and their uncertainties. For example, the temperature scale of the models may be too cool so that the observed temperatures and luminosities are only/mostly matched for a wrong mass.   

Five of the six target giants have been observed by the \textit{Kepler} mission, but only two have asteroseismic parameters presented in the literature; KIC8461659 is in the APOKASC catalogue \citep{Pinsonneault2018}, KIC\,8329894 was studied by \citet{Balona2013} while both have average asteroseismic values published by \citet{Yu2018}. We give their asteroseismic values in Table~\ref{tab:data}. Since their values are compatible with ours, the corresponding asteroseismic masses would also be in agreement with our results.

Five of the stars have masses estimated by "the cannon" \citep{Ness2016}, which estimates stellar mass from APOGEE spectra using an artificial neural network trained on the APOKASC catalogue by matching APOGEE spectra to masses based on asteroseismic scaling relations (without corrections). The similarity of their predicted masses and ours for the NGC\,6866 members, which we show in Table~\ref{tab:data},  suggests that "the cannon" works well for stars of this mass. However, given that there are not many stars of this mass in the APOKASC catalogue used for training, it should be verified that this is not due to an edge effect causing the machine to assign very similar masses to all stars with this type of APOGEE spectrum. The fact that "the cannon" assigns completely identical $T_{\rm eff}$ values to three of the stars could be hinting at such potential issues.

\citet{Leung2023} derived ages for stars with APOGEE spectra through a machine learning technique using asteroseismic ages from \citet{Miglio2021} as training input. While the idea is similar to "the cannon", the machine learning method is quite different. Their ages for our NGC6866 targets are given in Table~\ref{tab:seispropdata}. As seen, their age for the evolved HeCB star KIC\,8461659 is much too high, while for the others the ages are higher than ours, and not compatible with our results within their $1\sigma$ uncertainty. At first sight, one might think that the machine learning method of \citet{Leung2023} has a problem, but we found that instead the problem likely lies with their training ages; we plotted all stars with a mass above 2.5 $M_{\odot}$ in the \citet{Miglio2021} sample that \citet{Leung2023} used as their training set in the mass-radius diagram with our own targets. We found that the masses and radii clustered in a very narrow range in both mass and radius and yet the assigned ages ranged from 0.5 to 0.8 Gyr. This reveals two causes for the discrepant ages of \citet{Leung2023}; first, the large age deviation of the evolved HeCB giant KIC\,8461659 is caused by the fact that there is not a single evolved HeCB star in the training set. Second, the scatter in the input training ages from 0.5 to 0.8 Gyr for stars that compared to each other and our sample in the mass-radius diagram should all have an age very close to 0.5 Gyr explains their larger ages. While the cause for the high age scatter among very similar stars in the \citet{Miglio2021} sample remains unclear it is likely related to multimodal posteriors caused by the non-linear behaviour of evolutionary tracks at these masses. We consider a more detailed investigation outside the scope of the present paper.

\section{Summary, conclusions, and outlook}
\label{sec:conclusions}

We identify 6 HeCB stars as members of NGC\,6866 according to similarities in their positions, proper motions, parallaxes, radial velocities, metallicities, [C/N] values, effective temperatures, and luminosities. For five of the stars, we measure asteroseismic parameters and derive masses and radii, and find evidence that they are all HeCB stars, which also supports a common origin of the stars.
Comparing all information to stellar-model isochrones we estimate an age of NGC\,6866 of $0.43\pm0.05$ Gyr including an uncertainty of 0.1 dex in metallicity while constraining convective-core overshoot to be $\beta=$ 0.1 or less. Rotational effects seems to have had a different influence on the evolution of the HeCB stars in NGC\,6866 than suggested by current 1D stellar models.
Comparisons to literature studies of NGC\,6866 and/or the same stars as our study uncovered potential biases in mass and age estimates, which may propagate to estimates of field stars through machine learning algorithms and/or abundance-age relations. This underlines the importance of expanding this type of study to more open clusters in order to improve models and methods for increased precision and accuracy of stellar age measurements.
All four clusters in the \textit{Kepler} field have now been extensively studied using asteroseismology. However, there are still un-utilised cluster data from K2 where seismic analysis can be carried out. Unfortunately, the large pixel scale and the relatively short time span of observations in many regions hinders asteroseismology of solar-like oscillators in clusters using the TESS mission \citep{Ricker2014}, and although less severe, the large pixels will also prevent the upcoming PLATO mission \citep{Rauer2014} from detailed asteroseismic studies in the crowded fields of open and globular clusters. Instead, the future missions STEP\footnote{ https://space.au.dk/the-space-research-hub/step/} and Haydn \citep{Miglio2021b}, that are in planning, will be ideal to extend asteroseismology of cluster stars to a larger scale, to ensure proper stellar age estimates of both cluster and field stars of all ages and metallicities in the future.

\begin{acknowledgements}
We thank the anonymous referee for useful comments and suggestions that helped improve the paper.\\

We thank Jørgen Christensen-Dalsgaard and Jacob Rørsted Mosumgaard at the Stellar Astrophysics Centre at Aarhus University for discussions on the current state of stellar models including 3D simulation results for variable mixing length and T-$\tau$ relations.\\

Based in part on observations made with the Nordic Optical Telescope, owned in collaboration by the University of Turku and Aarhus University, and operated jointly by Aarhus University, the University of Turku and the University of Oslo, representing Denmark, Finland and Norway, the University of Iceland and Stockholm University at the Observatorio del Roque de los Muchachos, La Palma, Spain, of the Instituto de Astrofisica de Canarias.\\

This work has made use of data from the European Space Agency (ESA) mission {\textit{Gaia}} (\url{https://www.cosmos.esa.int/Gaia}), processed by the {\textit{Gaia}} Data Processing and Analysis Consortium (DPAC,\url{https://www.cosmos.esa.int/web/Gaia/dpac/consortium}). Funding for the DPAC has been provided by national institutions, in particular the institutions participating in the {\textit{Gaia}} Multilateral Agreement.\\
This paper includes data collected by the Kepler mission. Funding for the Kepler mission is provided by the NASA Science Mission directorate. Some of the data presented in this paper were obtained from the Mikulski Archive for Space Telescopes (MAST). STScI is operated by the Association of Universities for Research in Astronomy, Inc., under NASA contract NAS5-26555. Support for MAST for non-HST data is provided by the
NASA Office of Space Science via grant NNX09AF08G and by other grants and contracts.\\
This research has made use of the SIMBAD database,
operated at CDS, Strasbourg, France\\
This research made use of Lightkurve, a Python package for Kepler and TESS data analysis (Lightkurve Collaboration, 2018).\\
Funding for the Stellar Astrophysics Centre was provided by The Danish National Research Foundation (Grant agreement no.: DNRF106).\\
AM, AS, EW, JM, KB, MT and VG acknowledge support from the ERC Consolidator Grant funding scheme (project ASTEROCHRONOMETRY, \url{https://www.asterochronometry.eu}, G.A. n. 772293).
D.S. is supported by the Australian Research Council (DP190100666). 
ELS acknowledges support from the US National Science Foundation under grant AAG 1817217.

\end{acknowledgements}

\bibliographystyle{aa} 
\bibliography{References-1} 

\appendix

\section{Information on potential power spectra contaminants}

Here, we briefly mention the potential contaminants that we tried to avoid or minimze in the power pectra of our targets by adjusting the aperture used for extracting the \textit{Kepler} light curves.
\\

\subsection{KIC8329894}

KIC8329908 = \textit{Gaia} DR3 2076066901750428032 is a relatively bright (\textit{Gaia} G = 14.217) and near neighbour (12" away) that could cause issues when using the pipeline masks.	
\\

\subsection{KIC8395903}

KIC8395890 = \textit{Gaia} DR3 2082072502978526592 is near neighbour with G = 14.77. No \textit{Kepler} observations exist for this star, but the \textit{Gaia} parallax and colour suggest that it is a relatively cool main sequence non-member star, and thus not expected to exhibit any significant variability. All other neighbours to KIC8395903 within 40" are much fainter (G > 16.6).	
\\

\subsection{KIC8264549}

KIC8264581 is the nearest neighbour to KIC8264549 with a centroid distance of 13.2" and $V = 13.5$.
More problematic is KIC8264588 at a separation of 20", a known $\delta$ Scuti and binary star \citep{Murphy2018}, which gives rise to a false peak at about 58\,$\mu$Hz in the power spectrum of KIC8264549. With our custom mask we were able to reduce this peak very significantly, but we were not able to avoid it completely, and its amplitude remained at the same level as the highest peaks from the solar-like oscillations. Although it is placed outside of the frequency range of the solar-like oscillations (70--120\,$\mu$Hz), we removed the peak in the data analysis, prior to the determination of the asteroseismic parameters. We did this in two different ways; by removing the peak and replacing it with the average level in the power spectrum itself, and by prewhitening it from the time series before calculating the power spectrum. The latter method removes the effects of sidelobes stemming from the spectral window function, however due to the near-continuous nature of the \textit{Kepler} light-curves, the spectral window function is in this case very clean. Based on these measures, we are confident that the presence of this false peak does not affect our asteroseismic analysis. There is also another star located within the same central pixel, but it is 5.54 mag fainter in G than the oscillating target, so it is unlikely that it is causing significant noise in the power spectrum.
\\

\subsection{KIC8264592}

KIC8264590 = \textit{Gaia} DR3 2075877098557453568 is within 21". It has $G = 13.66$ which is about 13\% of the light of the target. 

\section{MCMC results visualised}

\begin{figure*}
   \centering
    \includegraphics[width=8cm]{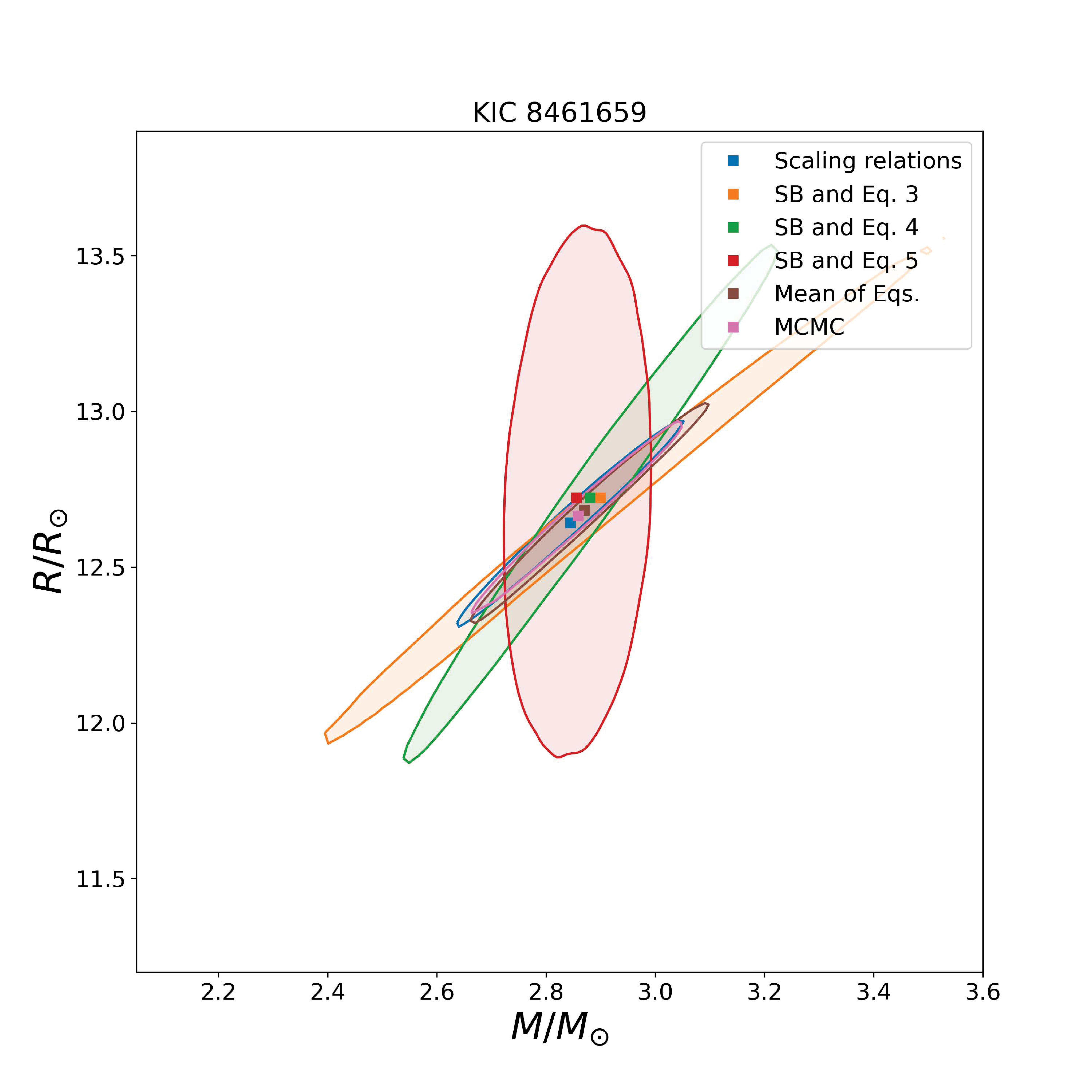}
    \includegraphics[width=8cm]{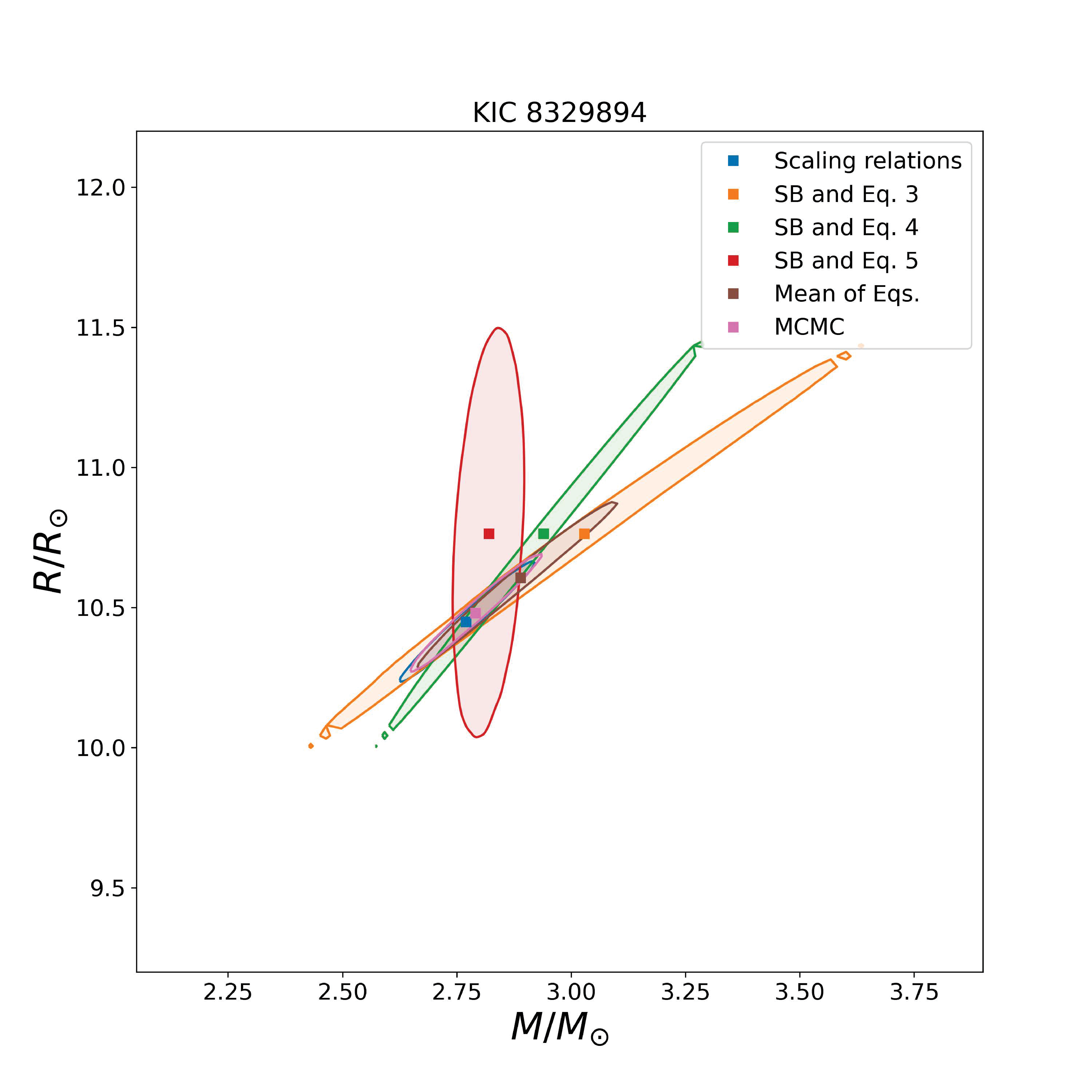}
    \includegraphics[width=8cm]{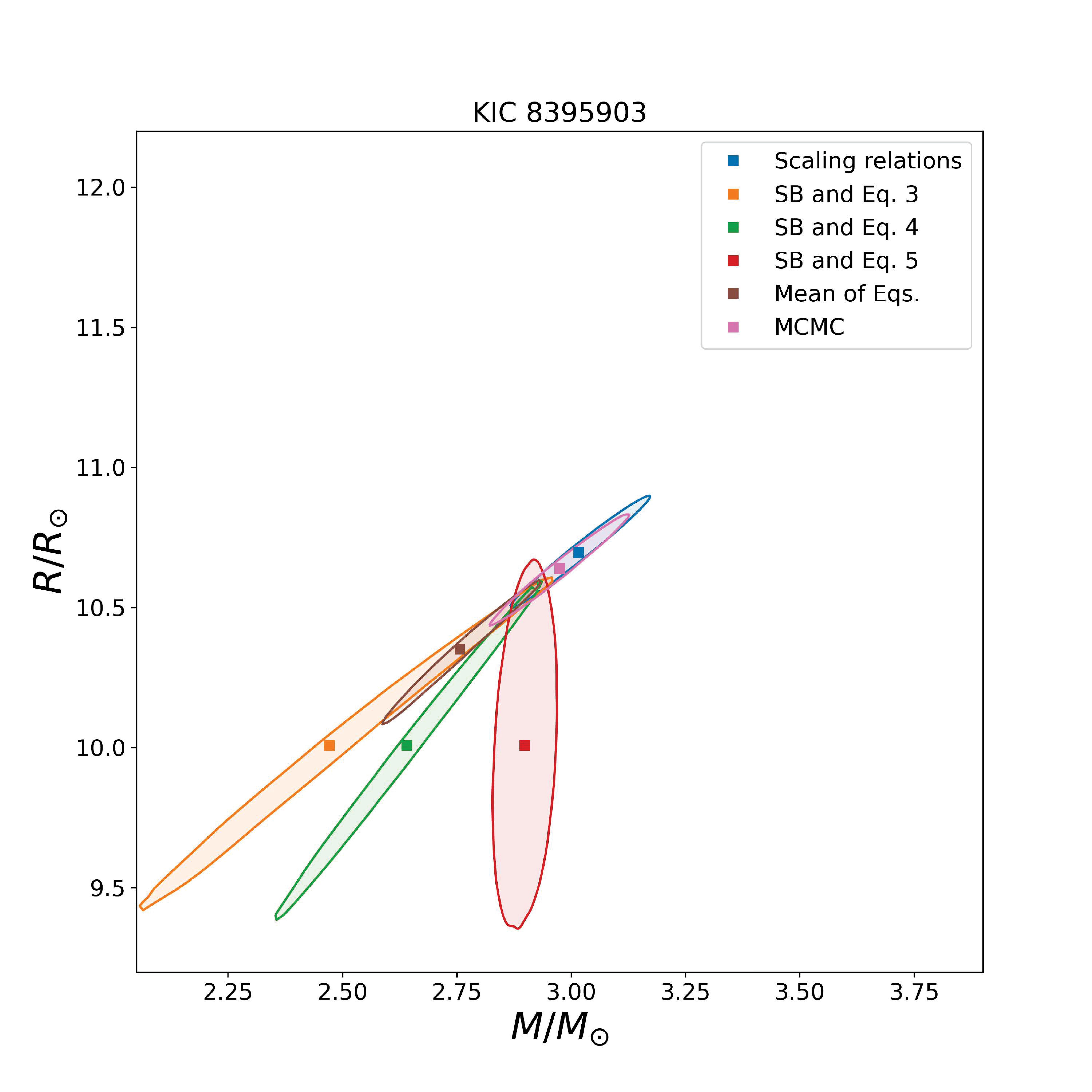}
    \includegraphics[width=8cm]{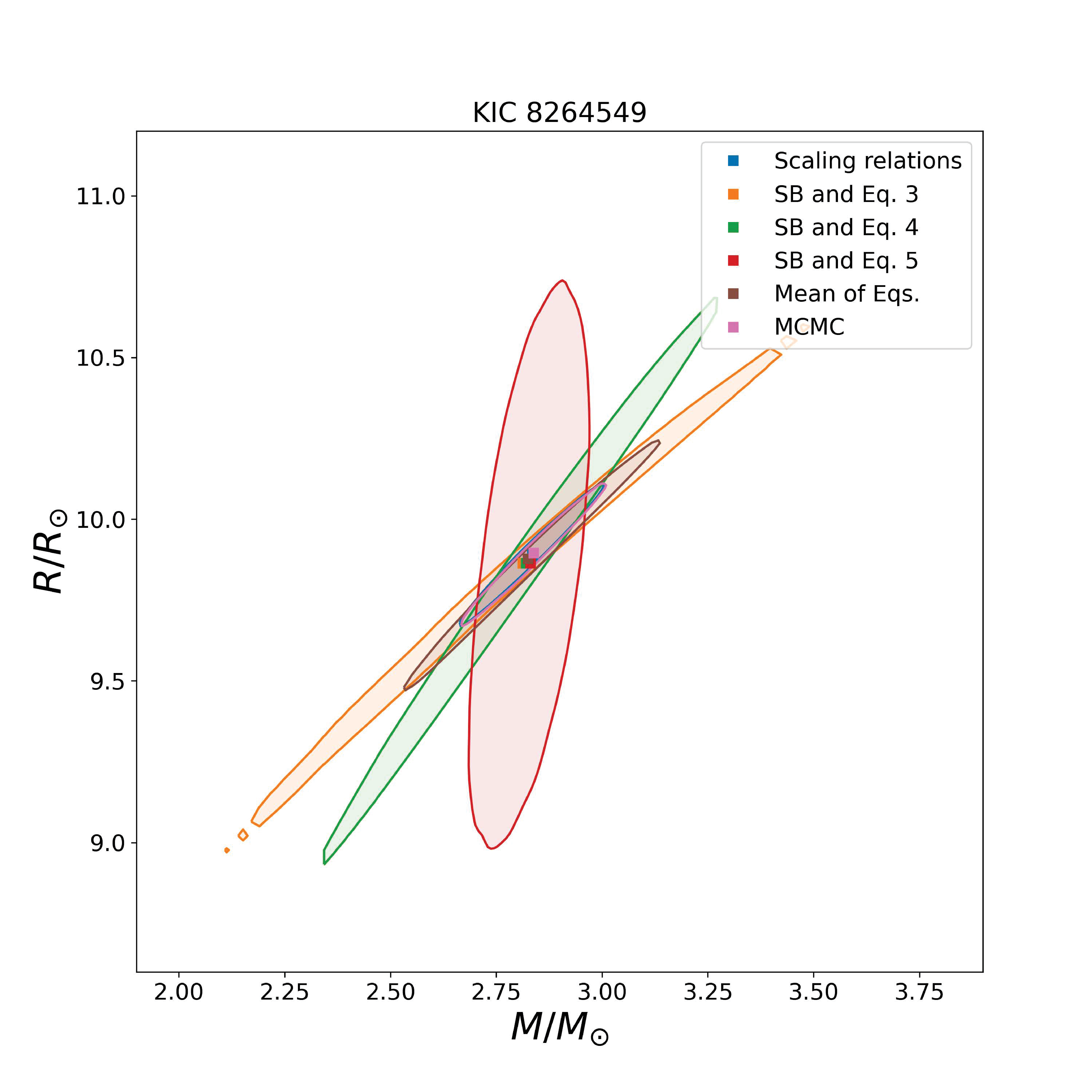}
    \includegraphics[width=8cm]{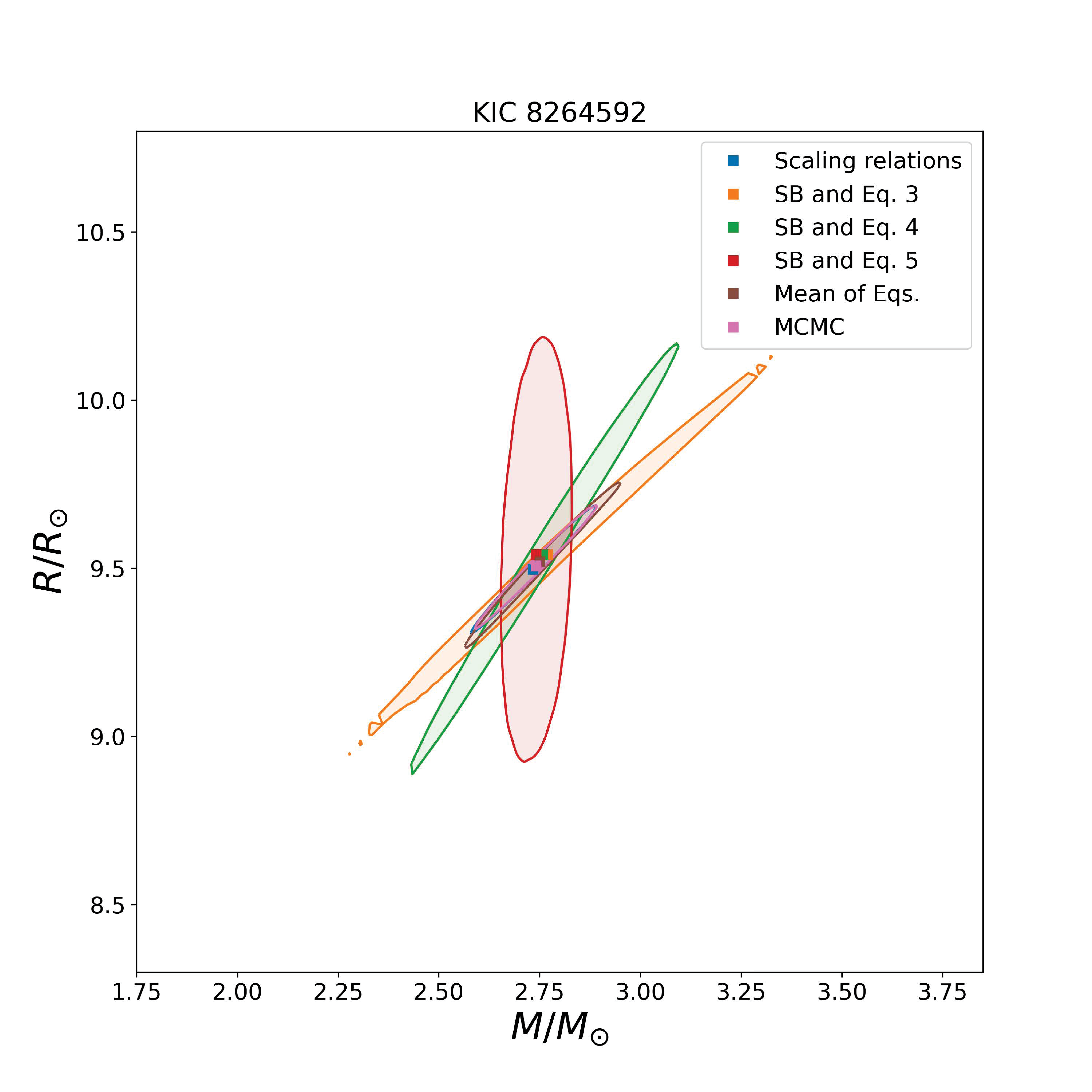}
      \caption{Mass-radius diagrams with 1$\sigma$ intervals for masses and radii calculated from different scaling relations and the Stefan-Boltzmann equation. The shaded contours mark the 1sigma confidence bounds, plotted using a Gaussian kernel density estimation. For all points except "MCMC", the samples were obtained with a standard Monte Carlo simulation procedure.}
         \label{fig:mcmc}
   \end{figure*}

\end{document}